\documentclass[showkeys,nofootinbib,longbibliography]{revtex4-2}

\usepackage[tbtags]{amsmath}
\usepackage{amssymb, amsthm}
\usepackage{physics, siunitx}
\usepackage{isomath}        
\usepackage{mathtools}      
\usepackage{subcaption}     
\usepackage{suffix}         
\usepackage{chemformula}    
\usepackage{multirow}
\usepackage{xspace}
\usepackage{comment}
\usepackage[T2A]{fontenc}
\usepackage[utf8]{inputenc}

\usepackage{algorithm}
\usepackage[noend]{algpseudocode}  
    \algnewcommand{\TRUE}{\textbf{true}}
    \algnewcommand{\FALSE}{\textbf{false}}
    \algnewcommand{\OR}{\textbf{ or }}
    \algnewcommand{\AND}{\textbf{ and }}
    \algnewcommand{\VAR}{\texttt}

\usepackage{titlesec}       
\usepackage{enumitem}       

\graphicspath{{figs/}}
\usetikzlibrary{arrows}

\usepackage[
    pdfencoding=unicode,
    psdextra,               
    colorlinks,
]{hyperref}
\hypersetup{
    colorlinks=true,
    citecolor = {blue}
}

\newcounter{RayTracing}

\newcommand{\tran}{\mathsf{T}} 

\DeclareSIUnit{\wtpercent}{wt\%}
\newcommand{\mysi}[1]{\si[per-mode=reciprocal]{#1}}

\NewDocumentCommand{\myoverbracetext}{m}{\text{#1}\\} 
\NewDocumentCommand\myoverbrace{ O{\int} m >{\SplitList{\\}}m }
{
    \overbrace{\vphantom{#1}#2}^{\substack{\ProcessList{#3}{\myoverbracetext}}}
}
\NewDocumentCommand{\myunderbracetext}{m}{\text{#1}\\} 
\NewDocumentCommand\myunderbrace{ O{\int} m >{\SplitList{\\}}m }
{
    \underbrace{\vphantom{#1}#2}_{\substack{\ProcessList{#3}{\myunderbracetext}}}
}

\newcommand\Dv[2][]{\frac{\mathrm{D}#1}{\mathrm{D}#2}}
\WithSuffix\newcommand\Dv*[2][]{\mathrm{D}#1/\mathrm{D}#2}

\newcommand{\gradf}[1]{\grad_{\!\! #1}}

\newcommand{\sol}{\text{S}}
\newcommand{\liq}{\text{L}}
\newcommand{\met}{\text{M}}
\newcommand{\gas}{\text{G}}

\newcommand{\melt}{\text{m}}
\newcommand{\laser}{\text{b}} 
\newcommand{\fus}{\text{fus}}
\newcommand{\newt}[1]{{#1}^{n+1}}
\newcommand{\oldt}[1]{{#1}^{n}}
\newcommand{\oldi}[1]{{#1}^{*}}
\newcommand{\zeri}[1]{{#1}^{0}}

\newcommand{\corr}{\text{corr}}
\newcommand{\sigm}{\mathcal{S}}

\newcommand{\bv}{\vb*{v}}
\newcommand{\bn}{\vu*{n}}

\newcommand{\btau}{\tensorsym{\tau}}

\newcommand{\pictsi}[1]{\si{#1}}
\newcommand{\axissi}[1]{[\si{#1}]}

\definecolor{orange}{rgb}{255,127,0}
\definecolor{violet}{rgb}{188,148,196}
\definecolor{green4}{rgb}{0,0.6,0}
\definecolor{auburn}{rgb}{0.43, 0.21, 0.1}
\newcommand\revA [1]{#1}
\newcommand\revB [1]{#1}
\newcommand\revC [1]{#1}
\newcommand\revALL [1]{#1}

\begin{document}

\title{The influence of volumetric shrinkage on the metal solidification process under localized energy deposition}

\author{Daniil V. Panov}
\email[]{daniil.panov@skoltech.ru}
\author{\mbox{Oleg A. Rogozin}}
\email[]{o.rogozin@skoltech.ru}
\author{Oleg V. Vasilyev}
\email[]{o.vasilyev@skoltech.ru}
\affiliation{Center for Materials Technologies, Skolkovo Institute of Science and Technology, Moscow, 121205, Russia}

\begin{abstract}

Accurate simulation of metal melting and solidification under localized energy deposition is crucial for the advancement of beam-based manufacturing technologies.
This study presents an extended multiphysics model \revALL{that addresses a critical limitation of prior approaches by incorporating volumetric changes from phase transitions and thermal expansion, in addition to capillary and thermocapillary effects.}
Validation against the benchmark problems---including a one-dimensional Stefan problem, two-dimensional solidification with free surface, and axisymmetric laser melting---demonstrates the high fidelity of the proposed model in describing melt-pool dynamics and free-surface evolution.
The numerical implementation features a novel mass-correction algorithm \revC{that reduces the mass conservation error by several orders of magnitude},
while a smoothed mushy-zone formulation \revALL{in the enthalpy method} mitigates the discretization \revC{artifacts in solid--liquid interface tracking}.
\revC{The results indicate that volumetric shrinkage plays an important role in} surface topography formation during solidification.
\end{abstract}


\maketitle

\section{Introduction}
\label{sec:introduction}
Spatially localized energy deposition that initiates metal melting and subsequent material redistribution is a fundamental thermophysical process underlying modern manufacturing technologies. In welding, a high-energy source is used to join parts together through targeted fusion~\cite{katayama2020fundamentals}. Additive manufacturing harnesses a focused beam to build three-dimensional structures layer by layer through selective melting of the feedstock material~\cite{Gibson2020AdditiveManufaturing}. In beam surface processing techniques, the redistribution of material within a shallow melt is precisely controlled to modify surface properties and topography~\cite{Earl2012surfisculpt,Temmler2012structuring}.

As these technologies advance and expand into new applications across various industries, the demand for in-depth understanding of the underlying physical mechanisms is constantly growing. This fundamental knowledge is essential for optimizing processes, ensuring quality control, and driving technological innovation. However, the experimental investigation of these physical mechanisms faces significant challenges. Conventional experimental techniques are limited to post-process characterization, offering  no detailed insights into the transient melt-pool dynamics. Although real-time monitoring techniques have been developed, they generally require substantial preparation and are typically customized for specific experimental setups~\cite{Cunningham2019Xray}.

These limitations have stimulated increasing adoption of numerical simulations as complementary research tools~\cite{DebRoy2018Review,Cook2020Review,Markl2016Review,DAL20162}. Significant advances have been achieved in understanding key mechanisms including: pore formation in additive manufacturing~\cite{Zhang2024porosity,Saad2020MeltPool,Khairallah2016modelingadditive}, the impact of laser beam shape on processing outcomes~\cite{Ebrahimi2023beamshape,Zhang2019dualbeam}, crystal growth patterns during solidification~\cite{Chen2016GrainStructure, Paul2020Dendrite,Yang2021phasefield}, and surface morphology evolution during melting~\cite{Ge2021surfacemodelling,Otto2015surfacemorphology}. Such simulations provide valuable insights that are often experimentally inaccessible.

The choice of a suitable mathematical model for numerical simulations largely depends on the specific research objectives. The spectrum of mathematical models used in simulation of additive manufacturing~\cite{BAYAT2021102278} and welding~\cite{DAL20162} vary in complexity from simplified heat-conduction equations  to sophisticated multi-physics formulations. The latter have been proven effective in accurately reproducing  melt-pool dynamics and resulting surface topography, which is crucial for studies of track morphology~\cite{Chen2018singlemorphology} and defect formation~\cite{Martin2019porosity}.

Advanced multi-physics models of metal melting processes integrate several interrelated physical phenomena, including heat transfer, phase transitions, surface tension, Marangoni convection, vaporization kinetics, beam optics, and beam--matter interactions~\cite{Zakirov2020EBMSimulation,Ma2024SphMelt}. Despite their complexity, these high-fidelity models share a fundamental simplifying assumption: they disregard volumetric changes resulting from both solid--liquid phase transitions and thermally-induced density variations in the molten phase. Instead, thermal expansion effects are typically approximated via the Boussinesq formulation~\cite[e.g.,][]{gutler:2013,panwisawas-2017}. While some earlier studies on phase change materials have addressed volumetric changes due to the phase transition~\cite[e.g.,][]{Chiang1992,Galione2015,Dallaire2017,Faden2019}, the influence of the phase change on the free surface is either omitted~\cite{Galione2015,Faden2019} or substantially simplified~\cite{Chiang1992,Dallaire2017}.  This omission is significant because volumetric shrinkage directly affects the final surface topography. Several studies have presented a systematic approach on how to incorporate a density jump in phase transition into the governing equations~\cite{Raessi2005,Thirumalaisamy2023}. However, to date, no studies have simultaneously accounted for volumetric changes resulting from both phase transitions and thermal expansion. Furthermore, the combined effects of capillarity, thermocapillarity, and density changes due to both phase transformation and thermal expansion have not been previously reported. Neglecting these coupled effects can lead to inaccuracies in predicting surface morphology.

Computational platforms for simulating melt-pool dynamics on fixed meshes---whether commercial or research-oriented---are based on a few established modeling approaches. Among them, the enthalpy-porosity method  is  the most prominent for describing thermo-fluid dynamics in solidification systems~\cite{1087EntVollerhalpyPorosity}. Free-surface boundary conditions are typically enforced using continuum surface force formulation~\cite{Brackbill1992CSF}, while evolution of a free surface is captured through Eulerian methods such as level set~\cite{Osher1988LevelSet} or volume of fluid~\cite{hirt1981volume}. Solution strategies commonly employ segregated schemes with pressure--velocity coupling~\cite{Issa1986}.
This pressure-based numerical formulation can be improved by incorporating density variations into the governing equations, thereby enhancing physical fidelity without the need for fundamental algorithmic modifications.

Studies utilizing the aforementioned numerical methodology to investigate the effects of volumetric changes can be found outside the field of beam-based metal manufacturing technologies~\cite[e.g.,][]{Koch2016bublecollapse,Miller2013UDE}.  Koch et al.~\cite{Koch2016bublecollapse} along with the work of Thirumalaisamy et al.~\cite{Thirumalaisamy2023}, have revealed a critical challenge in maintaining global mass conservation, showing that mass conservation errors can be comparable to the density variations themselves. Notably, Koch et al.~\cite{Koch2016bublecollapse} observed mass increases of up to 100\% in bubble cavitation simulations. Although the importance of mass conservation is widely recognized, the issue remains unresolved. For example, Miller et al.~\cite{Miller2013UDE} acknowledge mass conservation concerns but provide no substantive analysis. Thirumalaisamy et al.~\cite{Thirumalaisamy2023} initially found that relative mass errors decrease with increasing mesh resolution; however, in their subsequent work~\cite{Thirumalaisamy2024}, it was shown that global mass conservation deteriorates with increasing mesh resolution. Koch et al.~\cite{Koch2016bublecollapse} use a  direct density adjustment method to enforce mass conservation, representing one potential solution to this persistent problem.

In the present study, global mass conservation is achieved indirectly by adjusting the dilatation rate, which, due to the ellipticity of the pressure equation, results in compensatory displacement of the liquid metal free surface.

The present study has two main objectives: (1) to develop a comprehensive framework for simulating melt-pool dynamics that incorporates the effects of volumetric changes; and (2) to demonstrate that the developed approach ensures global mass conservation throughout the entire melting-solidification cycle.

The rest of the paper is organized as follows. Section~\ref{sec:assumptions} introduces the primary physical assumptions underlying the proposed model. The mathematical formulation is detailed in Section~\ref{sec:equations}. Section~\ref{sec:numerical} describes the numerical methodology, with particular emphasis on the proposed mass-correction procedure. Detailed numerical examples are provided in Section~\ref{sec:results} to highlight the significant impact of volumetric changes on free-surface dynamics and to verify global mass conservation. Finally, concluding remarks are presented in Section~\ref{sec:conclusions}.

\section{Key physical mechanisms and model assumptions}
\label{sec:assumptions}
The key physical mechanisms  and primary assumptions included in the proposed model are outlined below, prioritized by their relative importance. This study focuses on metal phase changes resulting from  surface energy deposition and the dynamics of the gas--metal interface considering three phases: liquid metal, solid metal and gas. \revALL{The gas--metal mixture is treated as immiscible, ensuring a well-defined gas--metal interface boundary. The phase transition in metal is assumed to occur isothermally within a mushy zone, with a density jump at the melting temperature. The metal density is assumed to be constant in solid state and temperature-dependent in the liquid state.} The characteristic spatial scales of the melt pool are substantially larger than the absorption depth of the incident beam, so the energy deposition is approximated as a surface heat source. The melt pool dynamics is strongly influenced by surface-tension related phenomena, therefore, both capillary and thermocapillary effects at the gas--melt interface \revALL{should be included in the model.} \revB{Evaporation effects are not addressed in this study, as the scope is limited to the conductive melting regime.} The liquid metal is treated as a Newtonian fluid. The wetting angle for the liquid metal on its own solid phase is assumed to be negligible, implying purely hydrophilic behavior. Both the specific heat capacity and thermal conductivity of the metal are assumed to vary linearly with temperature.

\revALL{A critical assumption of the proposed model  concerns the coexistence of non-deformable solid and thermally dilated liquid metal phases. Since the acoustic time scales in the liquid metal are negligible compared to characteristic time scales of the processes under consideration, the low-Mach number approximation~\cite{Rehm1978, Paolucci1982, Majda1985} is employed, where pressure is decomposed into a spatially constant thermodynamic pressure and a locally varying dynamic pressure, leading to a compressible system where density variations are driven solely  by temperature changes and phase transitions, while the dynamic pressure acts globally and instantaneously, similar to an incompressible case. Since the problems of interest are characterized by the presence of open boundaries and/or a free surface, the thermodynamic pressure is assumed to be constant in time. Given the simplified nature of the model, which treats the solid metal as a non-deformable medium with constant density, the variable density of the liquid metal can only be physically accommodated by the presence of a free surface or an open boundary. Otherwise, if the liquid phase is fully confined by solid metal, the thermal expansion or contraction of the liquid would lead to unphysical scenarios due to the fundamental incompatibility between the need to accommodate the volume change in the liquid phase and the constant-volume constraint imposed by the rigid solid surrounding it. Modeling such scenarios would require further sophistication of the model, such as by introducing compressibility or deformability in the solid phase.}

The dynamics of the gas phase is considered to be of minimal importance, primarily due to the negligible influence of the gas on the deformation of the liquid metal surface. Therefore, a simplified description of the gas phase is used: the gas is modeled as a Newtonian fluid with constant density, viscosity, specific heat capacity, and thermal conductivity. Due to this simplification and the limited impact of gas dynamics on the liquid metal flow, the gas flow is treated as largely fictitious, primarily serving to describe the gas--metal interface in an Eulerian specification.

\section{Mathematical model}
\label{sec:equations}
The mathematical model is developed for a fixed physical domain, denoted as $\Omega$, which includes  three distinct phases: gas, liquid metal, and solid metal. Each phase is defined by two primary variables: the metal volume fraction $\alpha \in [0,1]$ and the liquid metal volume fraction $\phi \in [0,\alpha]$. Here,  $\alpha = 0$ signifies the gas phase, while $\alpha = 1 $ indicates the metal phase. Within the metal phase, $\phi =0$ corresponds to the solid region and $\phi = 1$ denotes the fully liquid region.  In the mushy phase transition region, the liquid volume fraction varies within $\phi \in (0,\alpha)$.

The density is expressed as a weighted mixture
\begin{equation}
    \rho = (1-\alpha) \rho_\gas + \phi \rho_\liq(T) + (\alpha - \phi) \rho_\sol,
    \label{eq:density}
\end{equation}
where $\rho_\gas$, $\rho_\liq$, and $\rho_\sol$ represent the densities of the gas, liquid metal, and solid metal, respectively. As specified in Section~\ref{sec:assumptions}, both the gas density $\rho_\gas$ and the solid metal density $\rho_\sol$ are assumed to be constant, while the density of the liquid metal $\rho_\liq$ is assumed to be a function of temperature $T$.

The specific enthalpy of the multiphase system is given by
\begin{equation}
    \rho h = (1 - \alpha) \rho_\gas h_\gas + \phi \rho_\liq h_\liq + (\alpha-\phi) \rho_\sol h_\sol ,
    \label{eq:whole_enthalpy}
\end{equation}
with specific enthalpies for the individual phases defined at temperature $T$ as follows:
\begin{align}
    h_\gas &= h_{\gas 0} +  \int_{T_\text{ref}}^T c_{p\gas} ( \xi) \dd\xi, \\
    h_\sol &=  h_{\sol 0} + \int_{T_\text{ref}}^T c_{p\sol} ( \xi) \dd\xi, \\
    h_\liq &= h_\sol(T_\melt) + \int_{T_\melt}^T c_{p\liq} (\xi) \dd\xi + h_\fus,
\end{align}
where $h_{\gas 0}$ and $h_{\sol 0}$ are the specific enthalpies at the reference temperature $T_\text{ref}$, $c_{p\text{P}}$ denotes the specific heat capacity for each  phase $\text{P} \in \left\{\gas, \liq, \sol \right\}$, and $h_\fus$ is the latent heat of fusion.
\revC{For $T_\text{ref} = 0$, the baseline enthalpies, $h_{\gas 0}$ and $h_{\sol 0}$, are set to zero.}

The solid phase can be modeled using Brinkman penalization~\cite{arquis-caltagirone:1984}, which is based on the Brinkman's extension~\cite{brinkman:1947a,brinkman:1947b} of Darcy's law~\cite{darcy:1856}. This approach was further extended by Voller and Prakash~\cite{1087EntVollerhalpyPorosity} to model the mushy zone region in the phase-transition problems. In the present work, the mushy zone is treated as a  porous medium with finite permeability, while the solid phase is modeled with much lower permeability.

Surface forces along the gas--melt interface are effectively modeled using a delta function defined across the normal direction of the interface boundary,  which theoretically corresponds to physically consistent jump conditions. In  numerical implementation, the gas--metal interphase is represented by a smoothed continuous function $\alpha \in (0,1)$ that is localized along the interface. The gradient  $\grad\alpha$ defines the normal direction pointing from the gas to the solid region, with  $\bn_\alpha \equiv \grad\alpha/|\grad\alpha|$ defining the unit normal vector across the interface. The function $\left| \nabla \alpha \right|$  serves as the thickened delta function in the normal direction, effectively replacing  the interface conditions with localized volume sources  distributed across the thin interface region \cite{Brackbill1992CSF}.

Considering the  assumptions outlined in Section~\ref{sec:assumptions}, along with Eqs.~(\ref{eq:density}-\ref{eq:whole_enthalpy}), the Brinkman penalization approximation of the solid phase, and the interface simplifications discussed above, the conservation equations for mass in the gas and metal phases, as well as for momentum and energy, are formulated as follows:
\begin{gather}
  \pdv{\alpha}{t} +  \div (\alpha \bv) = \div \bv,
  \label{eq:advection_VOF}\\
 \div \bv = - \frac{(\rho_\liq  - \rho_\sol)D_t \phi + \phi \dv{\rho_\liq}{T} D_t T}{\phi \rho_\liq + (1-\phi) \rho_\sol},
  \label{eq:mass_conservation}\\
  \begin{multlined}
      \pdv{\rho\bv}{t}
      + \div (\rho \bv \otimes \bv)
      =
      - \grad p
      + \grad \vdot \btau + \myunderbrace{\rho \vb{g}}{gravity}
      - \myunderbrace{K\alpha\frac{(\alpha-\phi)^2}{\phi^2 + \epsilon}\bv}{Brinkman penalization}  \\
      - \myunderbrace{\gamma\qty(\div\bn_\alpha)\grad\alpha}{surface tension}
      + \myunderbrace{\dv{\gamma}{T}\qty(\bn_\alpha\cp\grad{T}\cp\bn_\alpha)|\grad \alpha|}{Marangoni force}\label{eq:momentum_conservation},
  \end{multlined}\\
      \pdv{\rho h}{t}
      + \div (\rho \bv h)
      = \ \div (k \grad{T})
      + \myunderbrace{I_{\laser} |\grad \alpha|}{beam heat source}, \label{eq:h_eq}
\end{gather}
where $\bv$ represents the velocity of the mixture, $p$ is the dynamic pressure,   $\btau = \mu [\grad \bv + (\grad \bv)^\tran ] - \frac{2}{3} \mu(\div \bv) \vb{I}$ is the viscous stress tensor, and $\vb{g}$ is the gravitational acceleration. The parameters $K$ and $\epsilon$ are used for Brinkman penalization, $\gamma$ denotes the surface tension, $k$ is the thermal conductivity, $I_{\laser}$ represents the  beam intensity, and $D_t \equiv \pdv*{t} + \bv\vdot\grad$ is the material derivative.
The dilatation rate, as defined by Eq.~\eqref{eq:mass_conservation}, consists of two terms: the phase-transition term, which is significant in the mushy region, and the thermal expansion term,  which accounts for the temperature-dependent density of the liquid metal.
Since the solid phase is modeled using Brinkman penalization, the same  viscosity is used for both the liquid and solid metal phases.   Thus, the viscosity  expressed as
\begin{equation}
    \mu = (1-\alpha) \mu_\gas + \alpha \mu_\liq,
    \label{eq:viscosity}
\end{equation}
where $\mu_\gas$ and $\mu_\liq$ are the viscosities of the gas and molten metal, respectively. Similarly, the thermal conductivity is defined as weighted average of the conductivities of different phases:
\begin{equation}\label{eq:conductivity}
    k = (1-\alpha)k_\gas + \phi k_\liq + (\alpha - \phi) k_\sol.
\end{equation}
Both Eqs.~\eqref{eq:advection_VOF} and \eqref{eq:mass_conservation} describe mass conservation in gas and metal phases, respectively.  Note that Eq.~\eqref{eq:mass_conservation} reduces to divergence-free condition in the gas and solid metal phases, whereas in the  liquid metal and mushy regions, it defines a non-zero dilatation rate.

The first under-braced source term in Eq.~\eqref{eq:momentum_conservation}  represents the Brinkman  penalization term. The  parameters $K$ and $\epsilon$ are selected such that permeability coefficient  in the solid metal phase is of order $\mathcal{O}(K/\rho_\sol \epsilon)$, resulting in an exponentially fast $\mathcal{O}(\exp(-Kt/\rho_\sol \epsilon))$ decay of velocity to zero. In  the mushy region, where liquid metal  volume fraction $\phi \in (0,\alpha)$, the permeability becomes of order  $\mathcal{O}(K/\rho_{\text{mushy}})$, with $\rho_{\text{mushy}}$ being the average density between the solid and liquid phases. The penalization term is automatically deactivated in both the liquid metal and gas phase regions. It should be noted that, in the original paper by Voller and Prakash~\cite{1087EntVollerhalpyPorosity}, $\epsilon$ is defined as a numerical parameter to avoid division by zero. However, in this study, $\epsilon$ is considered a parameter related to the time scale of velocity decay to zero in the solid region.

The other two under-braced source terms in Eq.~\eqref{eq:momentum_conservation} relate to surface tension forces as described earlier. The first term corresponds to the surface tension force, with the curvature of the gas--metal interface calculated as $\kappa = - \div \bn_\alpha$ (for details, see Ref.~\cite{enright:2002}). The second term represents the Marangoni force, which acts tangentially to the gas--metal interface. The tangential component of the temperature gradient $\div T$ is calculated as $\bn_\alpha \times \grad T \times \bn_\alpha$. Note that the deactivation of surface tension forces at the gas-solid metal interface is automatically achieved through the penalization term, which forces the velocity in the solid region to zero.
Similarly, the under-braced source term in Eq.~\eqref{eq:h_eq} corresponds to the surface heat term, as discussed in Section~\ref{sec:assumptions}.

The energy balance equation is formulated using an pure enthalpy formulation under the assumption of isothermal phase change, similar to the approach described in Ref.~\cite{fedorenko1975difference}. Unlike the standard enthalpy formulation found in Ref.~\cite{Voller1992GeneralEnthalpy},  the liquid phase fraction $\phi$ and the temperature gradient $\grad T$ are determined in terms of enthalpy in this paper.
The temperature $T$ is also derived from Eq.~\eqref{eq:whole_enthalpy}, along with assumptions about the specific heat capacities of the considered phases. The temperature gradient $\gradient T$ in Eq.~\eqref{eq:h_eq} is expressed in terms of specific enthalpy by differentiating Eq.~\eqref{eq:whole_enthalpy} as follows:

\begin{equation}
  \grad T = \frac{ \rho \grad h
  - (\rho_\sol( h - h_\sol ) - \rho_\liq(h - h_\liq) )\grad\phi
  - ( \rho_\gas( h - h_\gas)-\rho_\sol( h - h_\sol) ) \grad \alpha}
  {\rho c_p + \phi \rho_\liq^\prime(h_\liq - h)},
  \label{eq:temperature_gradient}
\end{equation}
where $\rho_\liq^\prime = \dv*{\rho_\liq}{T}$, and the mixture heat capacity $c_p$ is defined by the following equation:
\begin{equation}\label{eq:heat_capacity}
  \rho c_p = (1-\alpha)\rho_\gas c_{p\gas} +   \phi \rho_\liq c_{p\liq} +  (\alpha-\phi) \rho_\sol c_{p\sol}.
\end{equation}

The liquid metal fraction can be explicitly evaluated from Eq.~\eqref{eq:whole_enthalpy} as a function of specific enthalpy $h$ and metal volume fraction $\alpha$:
\begin{equation}
    \phi =
    \begin{cases}
        \begin{aligned}
        &0,  &h \leq h_1, \\
        &\frac{\alpha\rho_1(h-h_1)}{\rho_1(h-h_1) + \rho_2(h_2-h)}, &h_1  < h < h_2, \\
        &\alpha, &h \geq  h_2,
        \end{aligned}
    \end{cases}
    \label{eq:phi}
\end{equation}
where
\begin{equation}\label{eq:rho12}
    \rho_1 = (1-\alpha)\rho_\gas + \alpha\rho_\sol,  \quad
    \rho_2 = (1-\alpha)\rho_\gas + \alpha\rho_\liq(T_\melt),
\end{equation}
and
\begin{equation}\label{eq:h12}
    \begin{gathered}
    \rho_1h_1 = (1-\alpha)\rho_\gas h_\gas(T_\melt) + \alpha\rho_\sol h_\sol(T_\melt), \\
    \rho_2h_2 = (1-\alpha)\rho_\gas h_\gas(T_\melt) + \alpha\rho_\liq(T_\melt)h_\liq(T_\melt).
    \end{gathered}
\end{equation}
Note that $\phi(h_\melt) = 1/2$, where
\begin{equation}\label{eq:h_m}
    h_\melt = \frac{h_1\rho_1 + h_2\rho_2}{\rho_1 + \rho_2}.
\end{equation}

\section{Numerical method}
\label{sec:numerical}
The governing differential equations are solved using a cell-centered finite-volume method implemented in OpenFOAM~\cite{Weller1998OpenFoam}, with the \texttt{interIsoFoam} solver serving as the basis for the implementation. The essential modifications to the algorithm are detailed in this section. The source code is available in the repository\footnote{https://github.com/Mygetsy/foam-metal-shrinkage}.

A time-marching scheme based on a segregated algorithm is applied to solve the system of governing equations. For illustration purposes, a simplified version of the time discretization, based on the backward Euler method, is discussed in the following section. In practice, similar higher-order implementations can be used.
The notations $\newt{\psi}, \oldi{\psi}, \oldt{\psi}, \zeri{\psi}$ denote the current value of $\psi$, the value from the previous iteration, the value from the previous time step, and the value at initial time step, respectively.

\subsection{Volume fraction equation}

The volume fraction equation~\eqref{eq:advection_VOF} is numerically integrated using the geometric Volume of Fluid (VoF) method~\cite{hirt1981volume}, specifically the \texttt{isoAdvector} scheme, which, compared to algebraic VoF, offers improved advection accuracy, interface sharpness, volume conservation, and boundedness~\cite{roenby2016computational, scheufler2019geomvof}. For accurate simulations of the phase-transition processes, maintaining interface sharpness and boundedness of the volume fraction are critical due to the strong interaction between the smeared volume fraction of gas and the Brinkman penalization.

In the original study, Roenby et al.~\cite{roenby2016computational} introduced conservative \emph{bounding} and non-conservative \emph{clipping} procedures to enforce strict boundedness of the volume fraction, i.e., $\alpha\in[0,1]$. Similarly, the \texttt{interIsoFoam} implementation includes a \emph{snapping} procedure that enforces $\alpha$ to be $0$ or $1$ when values approach these bounds within a specified tolerance.
These necessary numerical interventions inevitably introduce mass conservation errors. According to Ref.~\cite{roenby2016computational}, non-conservative bounding is rarely required for incompressible cases. However, in the variable-density case,  the presence of a nonzero dilatation rate on the right-hand side of Eq.~\eqref{eq:advection_VOF} makes  these procedures essential to prevent unphysical values of  $\alpha$. The approach proposed to mitigate these mass conservation errors is presented in Sec.~\ref{subseq:mass_conserv}.

\subsection{Energy equation}

Given the updated volume fraction $\newt\alpha$, the discretization of the energy equation~\eqref{eq:h_eq} for specific enthalpy $h$ is expressed as follows:
\begin{eqnarray}\label{eq:h_scheme}
    \begin{split}
     \frac{\oldi{\rho}\newt{h} - \oldt{\rho}\oldt{h}}{\Delta t}
    &+ \div(\oldi{\rho}\oldt{\bv}\newt{h})
    = \div\frac{\oldi{\rho} \oldi{k}\grad{\newt{h}}}{\oldi{\rho} \oldi{c}_p + \oldi{\rho_\liq^\prime\!} (\oldi{h}_\liq - \oldi{h})}\\
    & - \div(\oldi{k} \frac{\rho_\sol(\oldi{h} - \oldi{h}_\sol ) - \oldi\rho_\liq(\oldi{h} - \oldi{h}_\liq)}{\oldi{\rho}\oldi{c}_p + \rho_\liq^{\prime\;*} (\oldi{h}_\liq - \oldi{h})}\grad\oldi\phi)\\
    & - \div(\oldi{k} \frac{
        \rho_\gas(\oldi{h} - \oldi{h}_\gas)-\rho_\sol(\oldi{h} - \oldi{h}_\sol)
      }{\oldi{\rho}\oldi{c}_p + \oldi{\rho_\liq^\prime\!} (\oldi{h}_\liq - \oldi{h})}\grad\newt\alpha)\\
    &+  I_\laser(\newt{t}) |\gradf{\oldi{\rho}\!\oldi{c}_p} \newt\alpha|.
    \end{split}
\end{eqnarray}
Hereafter, the notation $\gradf{f}\alpha$ is used to denote $\grad\alpha$, rescaled at the interface as
\begin{equation}
    \gradf{f}\alpha = \frac{2f(\alpha, T)}{f_\gas(0) + \phi_\met f_\liq(1,T) + (1-\phi_\met)f_\sol(1,T)}\grad\alpha,
\end{equation}
where $\phi_\met$ is the liquid fraction in metal, defined in the region $\alpha>0$ such that
\begin{equation}\label{eq:phi_met}
    \phi = \alpha \phi_\met.
\end{equation}
This approach, originally introduced by Brackbill et al.~\cite{Brackbill1992CSF} to enforce density-independent acceleration due to surface tension in the momentum equation, is employed here  to prevent spurious overheating of the gas phase resulting from interface thickening.

\subsection{Pressure-velocity coupling}

After obtaining  the solution for the energy equation, the coupled system of the momentum~\eqref{eq:momentum_conservation} and the dilatation rate~\eqref{eq:mass_conservation} equations is solved using a projection method similar to the PISO algorithm~\cite{Issa1986}. However, unlike the standard PISO algorithm, the projection step enforces the dilatation rate constraint~\eqref{eq:mass_conservation} to account for solid--liquid density jump and thermal density variations, rather than ensuring the solenoidality of the velocity field.
The momentum equation is discretized as follows:
\begin{equation}\label{eq:v_scheme}
    \begin{multlined}
    \frac{\newt{\rho}\newt{\bv} - \oldt{\rho}\oldt{\bv}}{\Delta t}
    + \div( \newt{\rho} \oldi{\bv} \otimes \newt{\bv} ) = -\grad \newt{p} \\
    + \div(\newt\mu \qty(\grad\newt{\bv} + (\grad \oldi{\bv})^\tran )) + \newt\rho \vb{g}
        - K \newt \alpha\frac{(\newt \alpha- \newt \phi)^2}{(\newt \phi)^2 + \epsilon}\newt{\bv}  \\
    - \newt \gamma(\div\bn_{\newt\alpha})\grad \newt\alpha
        + \dv{\newt{\gamma}}{T} \qty(\bn_{\newt\alpha}\cp\grad{\newt T}\cp\bn_{\newt\alpha}) |\gradf{\newt\rho} \newt\alpha|,
    \end{multlined}
\end{equation}
which can be rewritten as
\begin{equation}
    \newt{\bv} = A^{-1}\qty(\newt{r} - H(\oldi{\bv}, \oldt{\bv}) + \grad \newt{p}),
    \label{eq:new_velocity}
\end{equation}
where $H(\oldi{\bv}, \oldt{\bv})$ consists of matrix coefficients for all neighboring cells multiplied by the corresponding velocities, and $\newt{r}$ represents the source terms.

Since the density of metal is explicitly defined by the volume fractions $\alpha$ and $\phi$, as well as the temperature $T$, the continuity equation is not directly integrated. Instead, it is used to define the cell-centered values of the dilatation rate, which calculated as
\begin{equation}\label{eq:div_vel}
    \newt{(\div\bv)} = - \frac{(\rho_{\liq\melt} - \rho_\sol)\oldi{D}_t\newt\phi
    + \newt\phi \newt{\rho_\liq^\prime\!} \oldi{D}_t \newt{T} }{\newt{\phi} \newt \rho_{\text{L}} + (1-\newt{\phi}) \rho_\sol},
\end{equation}
where $\rho_{\liq\melt} = \rho_\liq(T_\melt)$ and
\begin{equation}
    \oldi{D}_t\newt\psi = \frac{\newt{\psi} - \oldt{\psi}}{\Delta t} + \oldi{\bv} \vdot \grad{\newt{\psi}}
\end{equation}
is the discretized form of the material derivative.
By taking the divergence of Eq.~\eqref{eq:new_velocity}, the variable-coefficient Poisson equation for the updated pressure takes the following form:
\begin{equation}
    \newt{(\div\bv)}  = \div(A^{-1}\qty(\newt{r} - H(\oldi{\bv}, \oldt{\bv}))) - \div(A^{-1} \grad \newt{p}).
    \label{eq:poisson_equation}
\end{equation}
The updated pressure $\newt{p}$ is obtained by solving Eq.~\eqref{eq:poisson_equation}. Subsequently, the velocity $\newt{\bv}$ is updated by substituting $\newt{p}$ into Eq.~\eqref{eq:new_velocity}. The left-hand side in Eq.~\eqref{eq:poisson_equation} is calculated using Eq.~\eqref{eq:div_vel}.

\subsection{Smoothing of liquid metal fraction}

\begin{figure}[!t]
    \centering

\begin{tikzpicture}[>=latex', scale=1.5]

    \def\xmax{3.4}
    \def\xstart{0.5}
    \def\ystart{0.5}

    \coordinate (A) at (\xstart,\ystart+0.25);
    \coordinate (L) at (\xstart+0.5,\ystart+1.25/2);
    \coordinate (LX) at (\xstart+0.5,0);
    \coordinate (LH) at (0,\ystart+1.25/2);

    \coordinate (B) at (\xstart+1,\ystart+1);
    \coordinate (BX) at (\xstart+1,0);
    \coordinate (BH) at (0,\ystart+1);

    \coordinate (M) at (\xstart+1.5,\ystart+1.5);
    \coordinate (MX) at (\xstart+1.5,0);
    \coordinate (MH) at (0,\ystart+1.5);

    \coordinate (C) at (\xstart+2,\ystart+2);
    \coordinate (CX) at (\xstart+2,0);
    \coordinate (CH) at (0,\ystart+2);

    \coordinate (R) at (\xstart+5/2*1.1,\ystart+4.5/2*1.05);
    \coordinate (RX) at (\xstart+5/2*1.1,0);
    \coordinate (RH) at (0,\ystart+4.5/2*1.05);
    \coordinate (D) at (\xstart+3,\ystart+2.5);

    \draw[->,thick] (0,0) -- (0,\xmax+0.2) node[left=0] {$h$};
    \draw[->,thick] (0,0) -- (\xmax+0.2,0) node[below=0] {$x$};

    \draw[thick,dashed] (A) -- (B);
    \draw[thick]        (B) -- (C);
    \draw[thick,dashed]        (C) -- (D);

    \draw[thick,dashed]        (L) -- (LX);
    \draw[thick,dashed]        (R) -- (RX);

    \draw[thick,dashed]        (B) -- (BX);
    \draw[thick,dashed]        (B) -- (BH);

    \draw[thick,dashed]        (M) -- (MX);
    \draw[thick,dashed]        (M) -- (MH);

    \draw[thick,dashed]        (C) -- (CX);
    \draw[thick,dashed]        (C) -- (CH);

    \draw[thick,dashed]        (L) -- (LH);
    \draw[thick,dashed]        (R) -- (RH);

    \fill[black] (B) circle(0.04);
    \fill[black] (BH) circle(0.04) node[fill=white,inner sep=0.5,left=2.5] {$h_1$};
    \fill[black] (L) circle(0.04)  node[inner sep=0.5,left=6.5, above=3.5] {$\phi = 0$};
    \fill[black] (LX) circle(0.04) node[fill=white,inner sep=0.5,below=2.5] {$x_l$};
    \fill[black] (LH) circle(0.04) node[fill=white,inner sep=0.5,left=2.5] {$h_l$};
    \fill[black] (R) circle(0.04)  node[inner sep=0.5,left=4.5, above=3.5] {$\phi = 1$};
    \fill[black] (M) circle(0.04);
    \fill[black] (MH) circle(0.04) node[fill=white,inner sep=0.5,left=2.5] {$h_m$};
    \fill[black] (RX) circle(0.04) node[fill=white,inner sep=0.5,below=2.5] {$x_r$};
    \fill[black] (RH) circle(0.04) node[fill=white,inner sep=0.5,left=2.5] {$h_r$};
    \fill[black] (C) circle(0.04);
    \fill[black] (CH) circle(0.04) node[fill=white,inner sep=0.5,left=2.5] {$h_2$};

\end{tikzpicture}
\caption{Specific enthalpy $h$ at computational nodes $x_l$, $x_r$ and the region between them, illustrating the liquid volume fraction $\phi$ when the mushy zone is located between the nodes. Quantities $h_1$ and $h_2$ are defined in Eq.~\eqref{eq:h12}, while $h_m$ is given by Eq.~\eqref{eq:h_m}.}
\label{fig:enthalpy-explanation}
\end{figure}
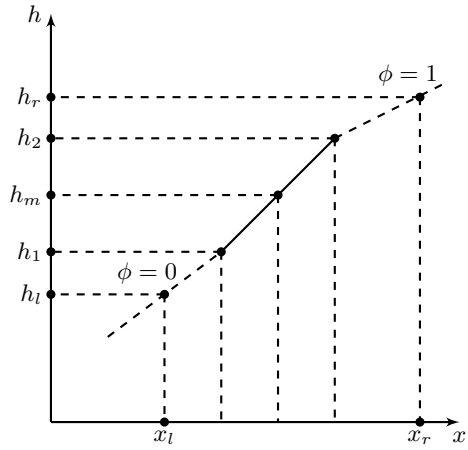

While the enthalpy formulation of the liquid phase volume fraction eliminates the issue of  multi-valued $\phi$ at the melting temperature, it introduces an essential numerical artifact: the \emph{stepwise} movement of the melt front. This oscillatory front propagation reduces accuracy of evaluating $\phi$, particularly when the mushy zone is positioned between the centers of computational nodes. As illustrated in Fig.~\ref{fig:enthalpy-explanation}, the computed values of $\phi$ at nodes $x_l$ and $x_r$ are 0 and 1, respectively, and remain unchanged until the enthalpy at one of the nodes reaches the mushy zone, where an abrupt change in the liquid fraction occurs, resulting in a non-smooth representation of the phase transition.

To mitigate this issue, the piecewise form of $\phi(h)$ defined by Eq.~\eqref{eq:phi} can be smoothed. A similar approach is also used in Refs.~\cite{Budak1965Smoothing, Azaiez2016SmoothingMelting}.
For this purpose, the volume liquid fraction in metal, $\phi_\met$, is approximated using a smooth sigmoid function
\begin{equation}\label{eq:sigmoid_def}
    \phi_\met(h) = \sigm\qty(\frac{\rho_1(h-h_1)}{\rho_1(h-h_1) + \rho_2(h_2-h)} - \frac12),
\end{equation}
where $\sigm(\xi) \in (0,1)$ is defined such that $\sigm(0) = 1/2$ and $\sigm'(0) = 1$.
Specifically, for numerical examples in this work, the error-function-based sigmoid function is used:
\begin{equation}\label{eq:erf_sigmoid}
    \sigm(\xi) = \frac12\erf\qty(\frac{\sqrt{\pi}}{4} (2\xi - 1)) + \frac12.
\end{equation}

\subsection{Mass correction}\label{subseq:mass_conserv}

The numerical implementation described in Section~\ref{sec:numerical} is inherently non-conservative~\cite{Thirumalaisamy2023,Miller2013UDE}. However, to accurately predict surface topography, ensuring global mass conservation in the metal phase is critical.
The discrepancy in mass conservation originates from
approximation errors in the dilatation rate determined by Eq.~\eqref{eq:div_vel}. Due to the elliptic nature of the pressure equation~\eqref{eq:poisson_equation}, these errors propagate over the entire melt pool,  ultimately resulting in inaccurate evolution of the gas-melt interface governed by the advection equation~\eqref{eq:advection_VOF}. Without mass conservation, the accumulation of these errors leads to a progressive deterioration of the mass balance,  necessitating the introduction of a mass-correction procedure.

To address the issue, the mass correction term is introduced into Eq.~\eqref{eq:div_vel}. In the absence of flow through the domain boundary, this correction term is modeled as linear feedback on the time scale $\tau$, proportional to the global mass imbalance between the current and initial time steps. The time scale $\tau$ governs the duration over which the mass correction is enforced. In the present study, $\tau$ is chosen to be equal to or greater than the time step. Thus, Eq.~\eqref{eq:div_vel} is reformulated using the definition of $\phi_\met$ from Eq.~\eqref{eq:phi_met} as follows:
\begin{equation}\label{eq:divU_corrected}
  \newt{(\div\bv)}  = - \newt\alpha\frac{(\rho_{\liq\melt}  - \rho_\sol)D_t^*\newt\phi_\met + \newt\phi_\met(\newt{\rho_\liq^\prime\!} D_t^* \newt{T} + \newt{\dot\rho}_\corr)}{\newt\phi_\met \newt\rho_\liq + (1-\newt\phi_\met) \rho_\sol},
\end{equation}
where
\begin{equation}\label{eq:rho_corr}
    \newt{\dot\rho}_\corr = \frac{\newt{m}_\met - \zeri{m}_\met}{\newt{V}_\liq \tau},
\end{equation}
and
\begin{equation}\label{eq:mass}
    m_\met(t) = \int \qty(\phi \rho_\liq + (\alpha - \phi)\rho_\sol) \dd{V}
\end{equation}
is the total mass of the metal at time step $t$, $\tau$ is the time scale of the mass-correction feedback, and $\newt{V}_\liq = \int_\Omega \newt\phi dV$ is the volume currently occupied by the molten metal. \revALL{Unlike direct mass-injection methods, such as that described in Ref.~\cite{Koch2016bublecollapse}, where the global mass conservation is enforced via artificial density adjustment, the present approach achieves conservation indirectly. Specifically, the dilatation rate equation~\eqref{eq:div_vel} is modified to incorporate a source term analogous to a uniform volumetric injection. In the presence of a free surface, this apparent volume change is immediately accommodated by adjusting the surface position during the pressure projection step using the Poisson equation~\eqref{eq:poisson_equation}, thereby ensuring global mass balance. Thus, the gas–melt interface is corrected through the advection step without modifying the density field.}

\subsection{Solution procedure}

The overall solution procedure is summarized in Algorithm~\ref{alg:pres}.
The system of discretized governing equations is numerically integrated in a segregated manner. At each time step, the algorithm executes at least two outer correction loops to ensure the coupled convergence of the primary variables. Within each outer loop, inner iterations are performed to sequentially solve the enthalpy equation~\eqref{eq:h_scheme} and the pressure equation~\eqref{eq:poisson_equation}. The convergence criteria for these inner iterations are determined by user-specified parameters.

\begin{algorithm}[H]
    \caption{Implemented solution procedure.}
    \label{alg:pres}
    \begin{algorithmic}
        \State Set initial conditions for $\alpha$, $\bv$, and $T$.
        \State Calculate $h^0$ and $\phi^0$ from $\alpha^0$ and $T^0$.
        \While{$\oldt{t}<t_\text{max}$}
            \State Update time $\newt{t} = \oldt{t} + \Delta{t}$.
            \While{not converged (outer correction loop)}
                \State Find $\alpha$ as a solution of Eq.~\eqref{eq:advection_VOF} using VoF.
                \State Calculate $\phi$ from $\alpha$ and ${h}$ using Eqs.~\eqref{eq:phi_met} and \eqref{eq:sigmoid_def}.
                \State Calculate $\rho$, $k$, and $c_p$ using Eqs.~\eqref{eq:density}, \eqref{eq:conductivity}, and~\eqref{eq:heat_capacity}.
                \While{not converged (enthalpy correction loop)}
                    \State Find $h$ as a solution of Eq.~\eqref{eq:h_eq} using scheme~\eqref{eq:h_scheme}.
                    \State Calculate $\phi$ from ${h}$ using Eqs.~\eqref{eq:phi_met} and \eqref{eq:sigmoid_def}.
                    \State Calculate ${T}$ from ${h}$ and $\phi$ using Eq.~\eqref{eq:whole_enthalpy}.
                    \State Calculate $\rho$, ${k}$, and ${c}_p$ from Eqs.~\eqref{eq:density}, \eqref{eq:conductivity}, and~\eqref{eq:heat_capacity}.
                \EndWhile
                \While{not converged (pressure correction loop)}
                    \State Calculate corrected $\div\bv$ using Eq.~\eqref{eq:divU_corrected}.
                    \State Find ${p}$ as a solution of Eq.~\eqref{eq:poisson_equation}.
                    \State Calculate $\bv$ from ${p}$ using Eq.~\eqref{eq:new_velocity}.
                \EndWhile
            \EndWhile
        \EndWhile
    \end{algorithmic}
\end{algorithm}

\section{Results and discussions}
\label{sec:results}
The capabilities and key features of the developed mathematical model and its numerical implementation are demonstrated in this section through the solution of several test problems, where density variations in metals play a significant role. The problems are presented in order of increasing physical complexity and variety of  physical mechanisms involved.

First, the relationship between melt front propagation and flow induced by volumetric changes is examined. Next, solidification in the presence of a free gas--metal interface is examined in a quasi-two-dimensional scenario involving flat free-surface solidification within a container subject to periodic boundary conditions. As the physical complexity increases, the coupled effects of capillarity and variable density are investigated in a two-dimensional solidification problem initiated from the side walls of the container. This is followed by a comparison with the classic benchmark problem of gallium melting in a rectangular two-dimensional container. Finally, an axisymmetric problem of metal melting  due to the energy deposition from an incident laser beam  is considered, incorporating all the phenomena assumed in the mathematical model.

\subsection{Thermophysical properties and model parameters}

Pure aluminum is chosen as the model material due to its relatively high density variation between the liquid and solid phases, with a thermal expansion coefficient comparable to other metals. Detailed thermophysical properties for the material are provided in Table~\ref{Tab:al_thermophysical}. The density of the liquid metal is assumed to vary linearly with temperature, as described by:
\begin{equation}
    \rho_\liq = \rho_{\liq\melt}(1 + \beta(T - T_\melt)),
\end{equation}
where $\rho_{\liq\melt}$ is the liquid metal density at the melting temperature and $\beta$ is the thermal expansion coefficient of the molten metal.

\begin{table}[t]
    \caption{Thermophysical properties of aluminum~\cite{mills2002recommended} and an auxiliary gas phases.}
    \centering
    \renewcommand{\arraystretch}{1.2}
    \scalebox{0.9}{
    \begin{tabular}{cccc}
        \hline
        Phase                                 & Physical property  & Value             & Unit                                 \\ \hline
        \multirow{5}{*}{Liquid metal}         & $\rho_{\liq\melt}$  & \num{2475}               & \mysi{\kg\per\cubic\m}          \\
                                              & $\beta$    & \num{e-4} & \mysi{\per\K}                       \\
                                              & $k_\liq$ & \num{91}                & \mysi{\W\per\m\per\K}  \\
                                              & $c_{p\liq}$     & \num{1042}              & \mysi  {\J\per\kg\per\K} \\
                                              & $\mu_\liq$    & \num{4.0e-5} & \mysi{\Pa\s}                        \\ \hline
        \multirow{3}{*}{Interface}   & $\gamma$     & \num{8.7e-1}    & \mysi  {\N\per\m} \\
                                              & $T_\melt$      & \num{936.6}             & \mysi{\per\K}    \\
                                              & $h_\fus$      & \num{3.8e5}             & \mysi{\J\per\kg}        \\ \hline
        \multirow{3}{*}{Solid metal}          & $\rho_\sol$   & \num{2700}              & \mysi{\kg\per\cubic\m}           \\
                                              & $k_\sol$ & \num{211}               & \mysi{\W\per\m\per\K}  \\
                                              & $c_{p\sol}$     & \num{910}               & \mysi{\J\per\kg\per\K} \\ \hline
        \multirow{4}{*}{Auxiliary gas phase} & $\rho_\gas$   & \num{4.0e-1} & \mysi{\kg\per\cubic\m}           \\
                                              & $\mu_\gas$    & \num{4.0e-5} & \mysi{\Pa\s}                        \\
                                              & $k_\gas$ & \num{6.1e-3} & \mysi{\W\per\m\per\K}  \\
                                              & $c_{p\gas}$     & \num{e3}              & \mysi{\J\per\kg\per\K} \\ \hline
        \end{tabular}
    }
    \label{Tab:al_thermophysical}
\end{table}

In addition to parameters obtained from experimental measurements, the model includes several parameters appearing in the penalization term of Eq.~\eqref{eq:momentum_conservation}, which that can be estimated.  As discussed in Sec.~\ref{sec:equations}, the permeability coefficient in the solid metal phase is typically of the order of $\mathcal{O}(K/\epsilon\rho_\sol)$, while in the mushy zone it is of the order of $\mathcal{O}(K/\rho_\text{mushy})$. To enforce the rigid behavior of the solid phase, the permeability of solid metal should be set as high as possible. Therefore, the parameter $\epsilon$ is chosen to be close to the machine precision limit as $\num{e-15}$. The coefficient $K$ for the mushy region is estimated using characteristic physical values. The characteristic velocity in the melt region is assumed to be $v_\text{melt} = \SI{e-1}{\m\per\s}$, the characteristic length scale of the mushy zone is taken as $l_\text{mushy} = \SI{5e-6}{\m}$, and the density is estimated to be $\rho_{\text{mushy}} = \SI{e3}{\kg\per\cubic\m}$. Based on these values, the coefficient $K$ is evaluated to be $K = \SI{5e7}{\kg\per\cubic\m\per\s}$.

\subsection{One-dimensional Stefan problem with variable density}

In two-phase solidification processes, density differences between phases induce flow within the liquid phase. This phenomenon is demonstrated through a one-dimensional Stefan problem, for which an analytical solution is provided in Refs.~\cite{Thirumalaisamy2023,Alexiades1993}. The following section highlights the ability of the proposed model to accurately capture the interface position, its relationship to the resulting flow in the liquid phase and the temperature distribution.

\subsubsection{Problem formulation}

The one-dimensional Stefan problem is formulated as follows. Initially, a large computational domain $\Omega \in [0,l]$, $l = \SI{1}{\m}$ is completely filled with liquid metal at a uniform temperature $2T_\melt$ and no flow at the initial stage, i.e.,
\begin{equation}
    T(x,0) = 2T_\melt, \quad v (x,0) = v_0,\quad p (x,0) = p_0,
\end{equation}
where $v_0 = \SI{0}{\m\per\s}, p_0 = \SI{0}{\Pa}$.
The boundary conditions are specified as follows:
\begin{eqnarray}
    \begin{split}
    T(0,t) &= 0.5T_\melt,\quad &T(l,t) &= 2T_\melt,\\
    v(0,t) &= v_0, &p(l,t) &= p_0.
    \end{split}
\end{eqnarray}

In this test case,  the density of the liquid metal density is assumed to be constant for comparison with the analytical solution. Consequently, the thermal expansion coefficient, $\beta$, is set to zero. Other thermophysical properties are provided in Table~\ref{Tab:al_thermophysical}.

\subsubsection{Results}

The relation between the induced by density differences and the melting front velocity is derived from the jump conditions at the interface and is expressed as follows:
\begin{align}
    \label{eq:1D_liq_vel_den_jump}
    u_\liq &= \qty(1 - \frac{\rho_\sol}{\rho_\liq})\dot{X}(t),\\
    \label{eq:1D_melt_interface_velocity}
    \dot{X}(t) &= \lambda\sqrt{\frac{k_\liq}{c_{p_\liq} \rho_\liq t}}\;,
\end{align}
where $\lambda$ is obtained from the similarity solution of the Stefan problem. At initial time, both the interface velocity $\dot{X}(t)$ and the liquid phase velocity $u_\liq$ become infinite, as seen in the solution~(\ref{eq:1D_liq_vel_den_jump}, \ref{eq:1D_melt_interface_velocity}). To exclude the influence of the initial time singularity on the numerical solution, the problem is solved starting from $t_{\text{s}} = \SI{100}{\s}$, using analytical solution as the initial condition.

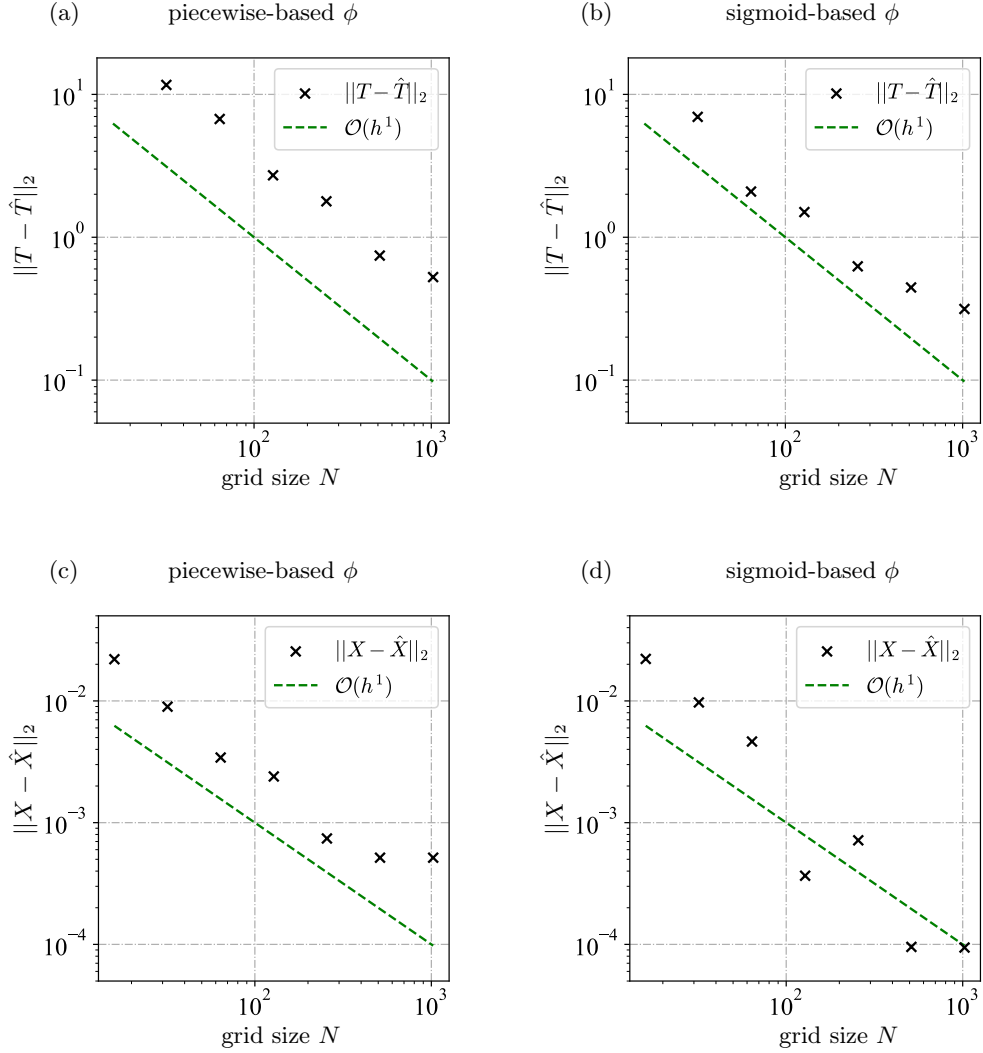
\begin{figure}[!t]
    \begin{picture}(400, 405)

    \put(0, 210){\includegraphics[width=0.33\textwidth]{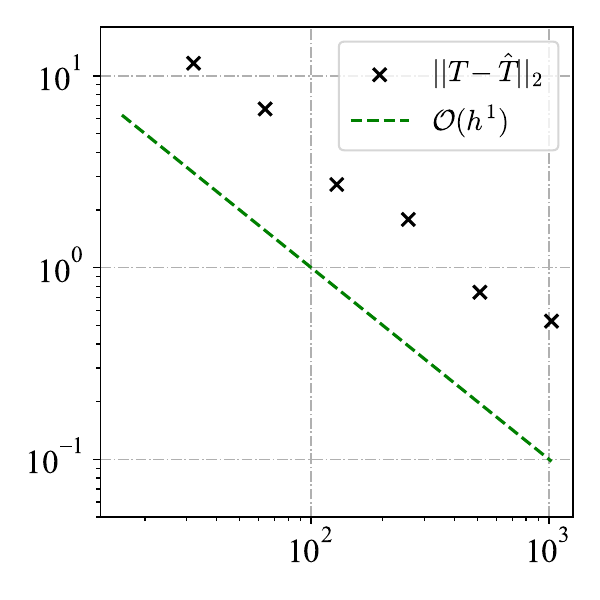}}
        \put(10,385){(a)}
    \put(55, 385){\small{piecewise-based $\phi$}}
    \put(75, 210){{grid size $N$}}
    \put(-5, 285){\rotatebox{90}{ $||T - \hat{T}||_2$}}

    \put(200, 210){\includegraphics[width=0.33\textwidth]{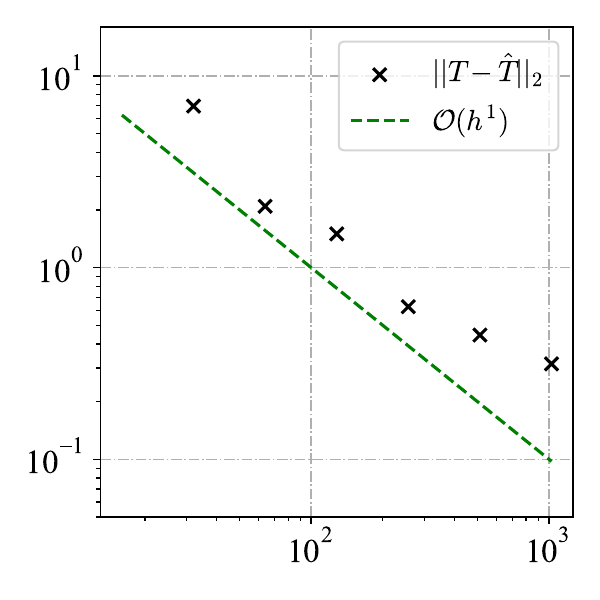}}
    \put(210,385){(b)}
    \put(265, 385){\small{sigmoid-based $\phi$}}
    \put(285, 210){{grid size $N$}}
    \put(195, 285){\rotatebox{90}{ $||T - \hat{T}||_2$}}

    \put(0, 0){\includegraphics[width=0.33\textwidth]{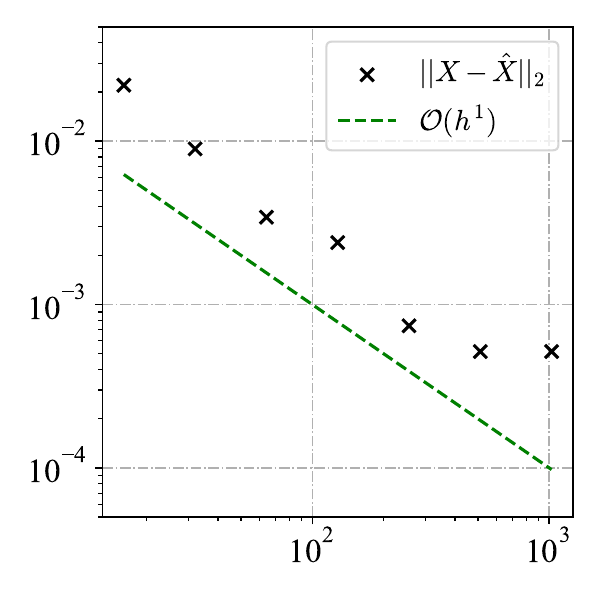}}
    \put(10,175){(c)}
    \put(55, 175){\small{piecewise-based $\phi$}}
    \put(75, 0){{grid size $N$}}
    \put(-5, 75){\rotatebox{90}{ $||X - \hat{X}||_2$}}

    \put(200, 0){\includegraphics[width=0.33\textwidth]{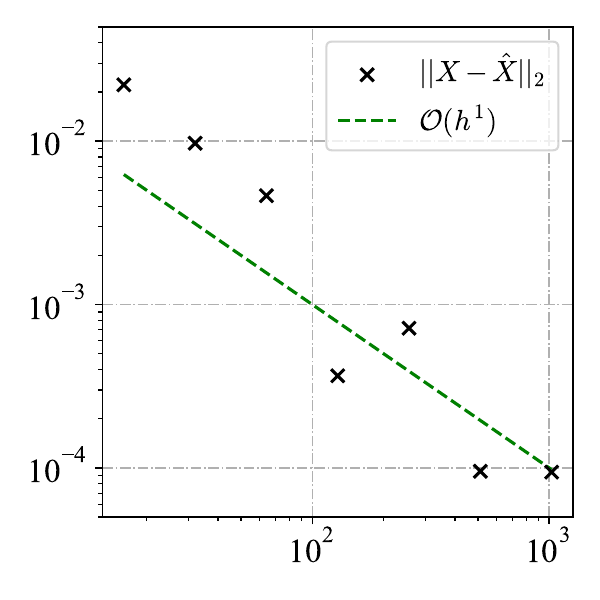}}
    \put(210,175){(d)}
    \put(265, 175){\small{sigmoid-based $\phi$}}
    \put(285, 0){{grid size $N$}}
    \put(195, 75){\rotatebox{90}{ $||X - \hat{X}||_2$}}

    \end{picture}

    \caption{Mesh convergence for the temperature distribution $T$ and RMSE for the melt front position $X$ across various grid sizes for $\phi_\met$, evaluated form the enthalpy $h$ using: (a, c) Eq.~\eqref{eq:phi_met} and (b, d) Eq.~\eqref{eq:sigmoid_def}.}
    \label{fig:T-X-grid-convergence}
\end{figure}

\begin{figure}[!t]
    \begin{picture}(400, 190)
        \put(0, 0){\includegraphics[width=0.33\textwidth]{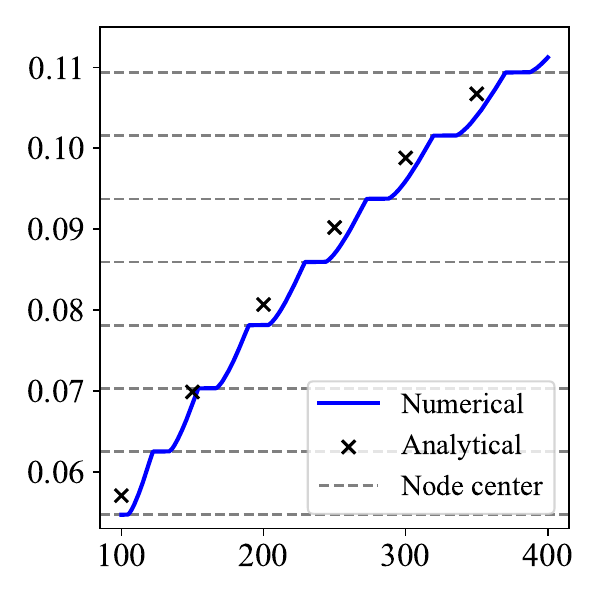}}
        \put(10, 175){(a)}
        \put(55, 175){\small{piecewise-based $\phi$}}
        \put(85, -2){{time $\axissi{\s}$}}
        \put(-5, 65){\rotatebox{90}{ position $X\; \axissi{m}$}}

        \put(200, 0){\includegraphics[width=0.33\textwidth]{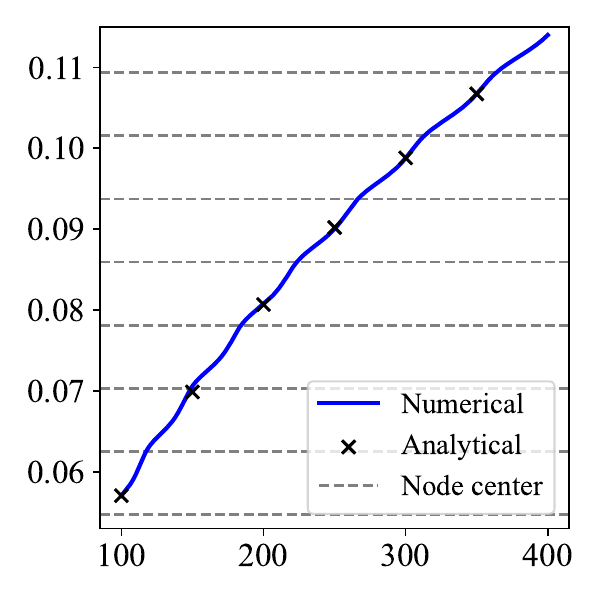}}
        \put(210, 175){(b)}
        \put(265, 175){\small{sigmoid-based $\phi$}}
        \put(285, -2){{time $\axissi{\s}$}}
        \put(195, 65){\rotatebox{90}{ position $X\; \axissi{m}$}}
    \end{picture}
    \caption{Comparison between the analytical melt front position $\hat{X}$ and the calculated position $X$  for a grid with $N=128$ nodes, evaluating  $\phi_\met$ from the enthalpy $h$ using: (a) Eq.~\eqref{eq:phi_met} and (b) Eq.~\eqref{eq:sigmoid_def}.}
    \label{fig:X_t_plots_with_mesh}
\end{figure}

Equation~\eqref{eq:1D_liq_vel_den_jump} illustrates the coupling between the movement of the melt front and the flow in the liquid phase. Therefore, it is important to obtain an accurate solution to the thermal problem in order to model the  flow in the liquid phase correctly. To assess the accuracy of the proposed numerical method for solving the one-dimensional Stefan problem, a grid convergence study is conducted. The number of grid points is varied as $N = \{16,32,64,128,256,512,1024\}$, while the time step is kept fixed and sufficiently small at $\Delta t = \SI{e-2}{\s}$ across all resolutions to ensure it does not affect spatial convergence. Convergence rates are examined for the temperature distribution and the melt front position. The temperature distribution $T$ is assessed  at $t=3t_{\text{s}}$, while  the spatial convergence of the melt front position is evaluated over the entire simulation duration. The error between the numerical solution $T$ and the analytical temperature distribution $\hat{T}$ is calculated using the discrete $L_2$-norm at grid points. The error between numerically obtained melt front position $X$ and the analytical position $\hat{X}$ is calculated  as the root mean square error (RMSE) over time steps.

The melt front position is computed from the distribution of the liquid volume fraction by integrating the distribution over the domain as follows:
\begin{equation}\label{eq:num_front}
    X(t) = \int_\Omega \phi\: dx,
\end{equation}
where the volume fraction $\phi$ is calculated form the enthalpy $h$ using either a piecewise function given by Eq.~\eqref{eq:phi} or a sigmoid-based on the error-function provided by Eq.~\eqref{eq:erf_sigmoid}.

The method demonstrates first-order convergence for  both the temperature distribution and the melt front position, as shown in Fig.~\ref{fig:T-X-grid-convergence}. A stepwise movement of the melt front is observed in Fig.~\ref{fig:X_t_plots_with_mesh}b, when  the piecewise liquid volume fraction  is used. This stepwise movement of the melt front is a well-known drawback of the enthalpy method, as reported in Refs.~\cite{Thirumalaisamy2023,fedorenko1975difference,Alexiades1993,Price1954StepHeat}. The effect of the stepwise movement becomes negligible at high grid resolutions. However, while the stepwise movement diminishes at finer grid resolutions, it remains significant in practical engineering applications, where computational constraints often necessitate the use of coarser meshes.

An example of the stepwise behavior can be observed in the analysis of the interphase location over time, as shown in Fig.~\ref{fig:X_t_plots_with_mesh}a. This stepwise movement is also reflected in the stepwise flow in the liquid phase via Eq.~\eqref{eq:1D_liq_vel_den_jump}. As demonstrated in Fig.~\ref{fig:X_t_plots_with_mesh}b, the use of sigmoid-based function to evaluate the liquid volume fraction mitigates this effect.

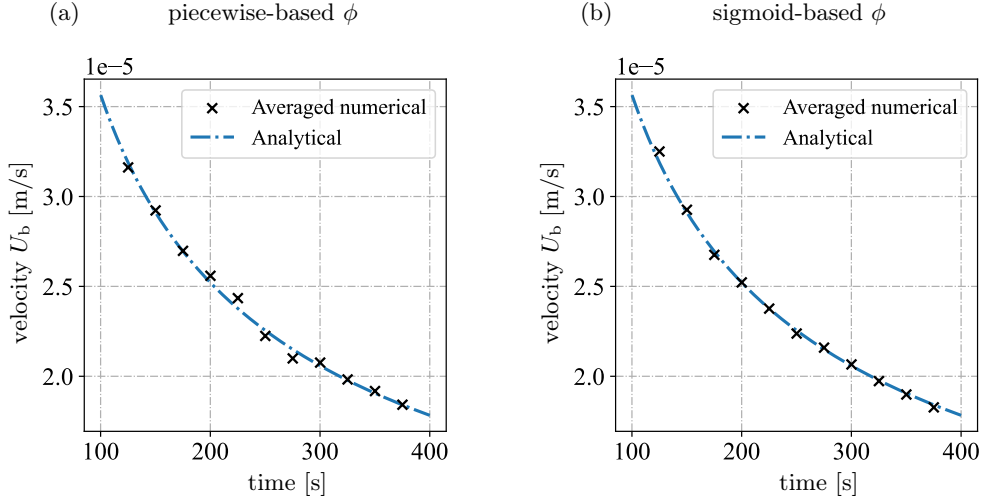
\begin{figure}[!t]
    \begin{picture}(400, 190)
        \put(0, 0){\includegraphics[width=0.33\textwidth]{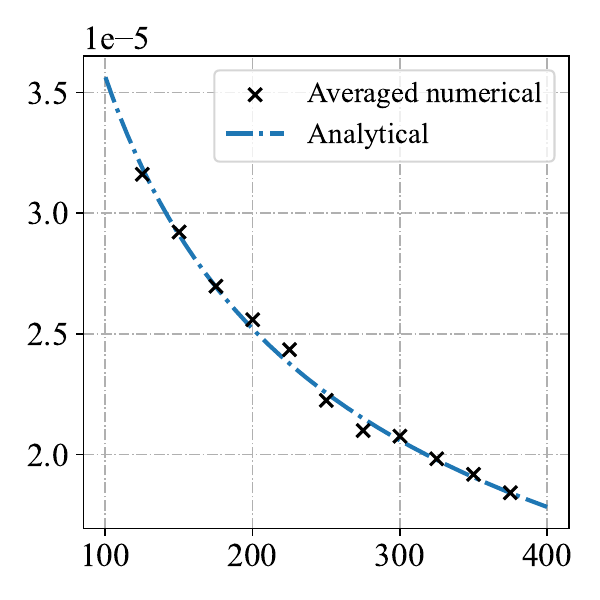}}
        \put(10, 175){(a)}
        \put(55, 175){\small{piecewise-based $\phi$}}
        \put(85, -2){{time $\axissi{s}$}}
        \put(-5, 50){\rotatebox{90}{ velocity $U_{\text{b}}\;\axissi{m/s}$}}

        \put(200, 0){\includegraphics[width=0.33\textwidth]{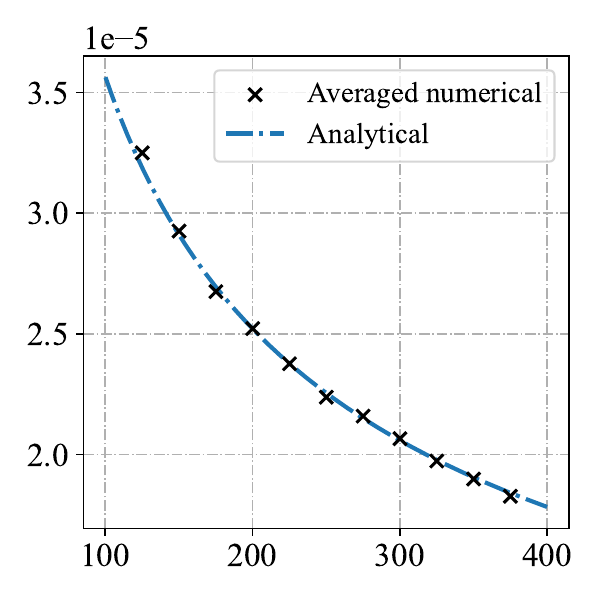}}
        \put(210, 175){(b)}
        \put(260, 175){\small{sigmoid-based $\phi$}}
        \put(285, -2){{time $\axissi{s}$}}
        \put(195, 50){\rotatebox{90}{ velocity $U_{\text{b}}\;\axissi{m/s}$}}
    \end{picture}
    \caption{Comparison between the  analytical solution $u_\liq$  and averaged calculated flow velocity $U_b$  at the outer boundary for a grid with $N=1024$ nodes, evaluating  $\phi_\met$ from the enthalpy $h$ using: (a) Eq.~\eqref{eq:phi_met} and (b) $\phi_\met$ from Eq.~\eqref{eq:sigmoid_def}.}
    \label{fig:U_t_plots}
\end{figure}

The effect of accounting for density change on the flow in the liquid phase is observed by analyzing the flow velocity at the outer boundary, $U_{\text{b}}$. Simulations were performed using a grid with $N=1024$ nodes. As seen in Fig.~\ref{fig:U_t_plots}, the calculated velocity at the outer boundary is in good agreement with the analytical solution for both the piecewise and sigmoid-based evaluations of $\phi_\met$. Although the stepwise  movement of the melt front introduces periodic deviations in the numerical solution,  the time-averaged velocity remains consistent with the analytical solution.

\subsection{2D planar solidification}

In the next test problem, a quasi two-dimensional solidification scenario  with a free gas--metal interface is considered. This problem is inspired by Refs.~\cite{Thirumalaisamy2023,Huang2022SolFree}, where the challenges of maintaining mass conservation during phase change processes are emphasized. In contrast to previous studies, in this work the additional effect of thermal expansion is taken into account. The problem is formulated to demonstrate the ability of the proposed model to describe phase transformations and interfacial motions in the presence of a free interface.

\subsubsection{Problem formulation}

The problem is defined in a rectangular domain $\Omega \in [0, l]\cross [0, l]$, where $l = \SI{8e-4}{\m}$. The thermophysical properties of aluminum are listed in Table~\ref{Tab:al_thermophysical}. Initially, the lower half of the domain is filled with liquid aluminum at a temperature twice the melting temperature, while the upper half contains gas at the same temperature. Both the liquid metal and gas start at rest, $|\bv_0| = \SI{0}{\m\per\s}$. This setup differs from Ref.~\cite{Thirumalaisamy2023}, as the gas temperature has been adjusted to achieve thermal equilibrium at the gas--metal interface at the onset of the simulations. The initial conditions are outlined below:
\begin{equation}
    T(x,y,0) = 2T_\melt,\quad \bv (x,y,0) = \bv_0.
\end{equation}

No-slip velocity boundary conditions are applied to the bottom  boundary of the domain, which is maintained at a fixed temperature of $0.5T_\melt$. The top boundary has adiabatic, stress-free, and open boundary conditions with a constant pressure. The domain is periodic in the lateral directions. The boundary conditions are summarized as follows:
\begin{align}
    \begin{split}
        T(x,0,t) &= 0.5T_\melt ,\quad {\pdv{T}{y}}(x,l,t) = 0,\quad T(0,y,t) = T(l,y,t),\\
        \bv(x,0,t) &= \bv_0,\quad \bv(0,y,t) = \bv(l,y,t),\quad p(x,l,t) = p_0.
    \end{split}
\end{align}

\subsubsection{Results}
\begin{figure}[!t]
    \centering

    \begin{picture}(400, 125)
    \put(-4, 0){
        \put(0, 10){\includegraphics[width=0.725\linewidth]{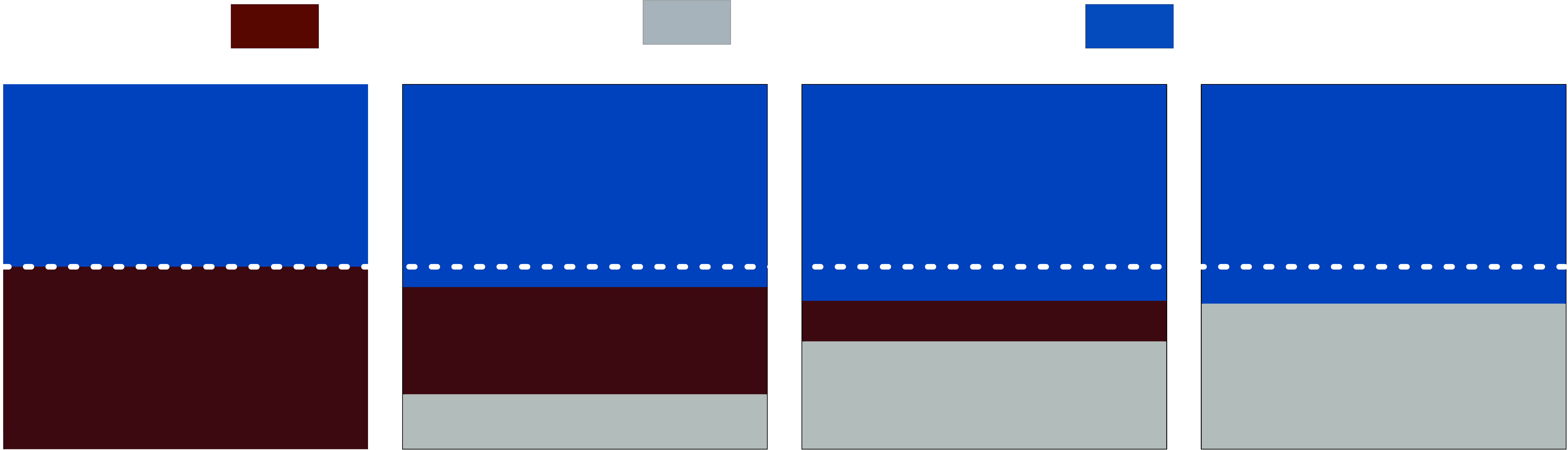}}
        \put(85, 108){
         -- \quad 
        \begin{tabular}{@{}c@{}} 
            molten \\ 
            metal     
        \end{tabular}%
    }
        \put(185, 108){ -- \quad
        \begin{tabular}{@{}c@{}} 
            solidified  \\ 
            metal     
        \end{tabular}%
        }
        \put(285, 108){ -- \quad gas}

        \put(45, 1){\makebox(0,0)[c]{ $t = 0\;\pictsi{s}$}}
        \put(143, 1){\makebox(0,0)[c]{ $t = 0.05\;\pictsi{s}$}}
        \put(235, 1){\makebox(0,0)[c]{ $t = 0.12\;\pictsi{s}$}}
        \put(331, 1){\makebox(0,0)[c]{ $t = 0.5\;\pictsi{s}$}}
        }
    \end{picture}

    \caption{\revA{Phase distribution in the horizontal solidification problem at different times and initial free surface position (white line).}}
    \label{fig:hor-case-solidif-viz}
\end{figure}
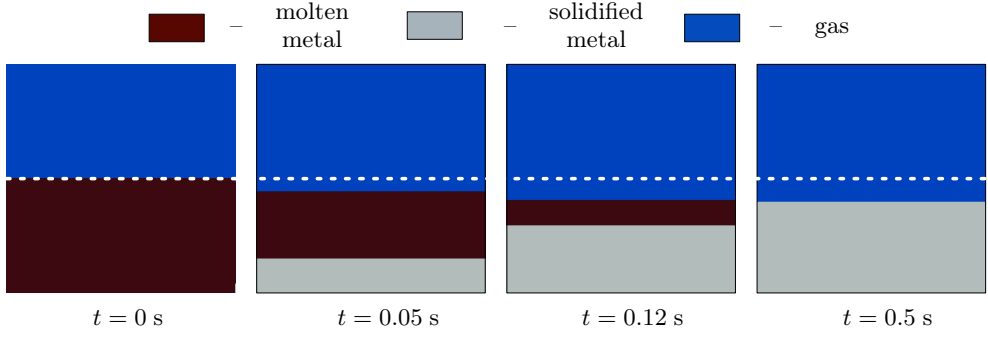

The volume change during the solidification process   is illustrated in Fig.~\ref{fig:hor-case-solidif-viz}, where the liquid and solid metal phases are shown at different stages of the numerical simulation. In the absence of metal flow through the walls, the mass of the metal within the domain should remain constant.  However, as demonstrated  in Refs.~\cite{Thirumalaisamy2023,Huang2022SolFree}, where a similar problem was considered, deviations from the initial mass were detected that were substantially larger than machine precision. Furthermore, mass error accumulation over time was observed in Ref.~\cite{Huang2022SolFree}. Additionally, it was shown in Ref.~\cite{Thirumalaisamy2023} that the mass error did not decrease with mesh refinement. This error accumulation could lead to significant deviations in the position of the gas--metal interface during long engineering calculations on a coarse grid. In the section, the non-conservative behavior is analyzed, and the effectiveness of the proposed mass correction in enforcing  mass conservation is demonstrated.

The mass conservation error is measured as the relative mass error in the domain, denoted as $e_m$, and is defined as follows:
\begin{equation}
    e_m(t) = \frac{\abs{\int\rho_\met(t)dV - \int\rho_\met(0)dV}}{\int\rho_\met(0)dV},
\end{equation}
where $\rho_\met(t) = \phi \rho_\liq + (\alpha - \phi)\rho_\sol$ represents the metal density at time $t$.
The relative mass error, $e_m(t)$, is examined throughout the entire computation time for various mesh configurations. The effect of mesh resolution is assessed using grids of size $N \times N$, where  $N$ takes values of 16, 32, 64, 128, and 256. The corresponding time steps, adjusted according to the mesh size,  are $10^{-4}$, $2.5\times10^{-5}$, $6.25\times10^{-6}$, $1.56\times10^{-6}$, and $3.9\times10^{-7}$. A piecewise liquid volume fraction is applied for all numerical experiments. Two cases are considered: the first without the application of non-conservative bounding for VOF (see Sec.~\ref{sec:numerical}), and the second with the application of clipping and snapping with tolerance of $\num{e-7}$. Since this problem serves solely as a numerical experiment to study mass conservation, the gravitational source term in Eq.~\eqref{eq:momentum_conservation} is neglected.

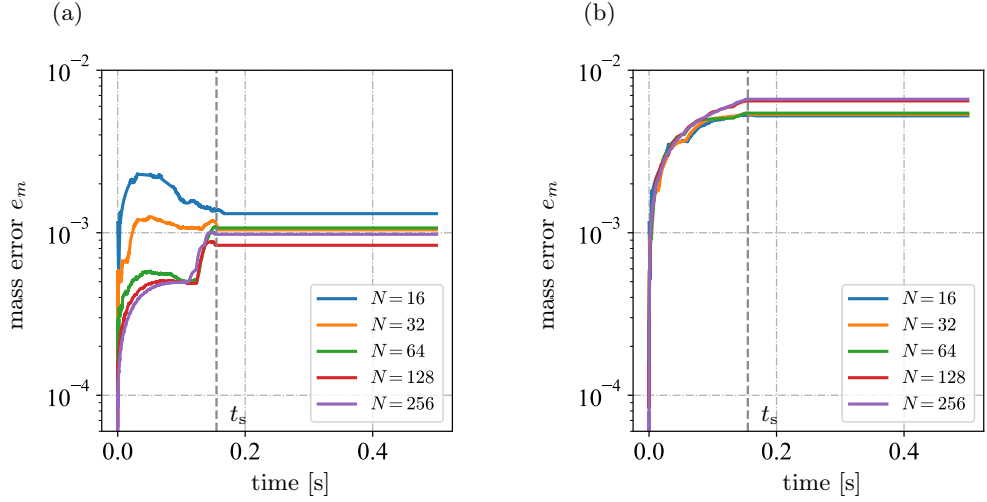
\begin{figure}[!t]
    \begin{picture}(400, 190)
    \put(5,0){
        \put(0, 0){\includegraphics[width=0.33\textwidth]{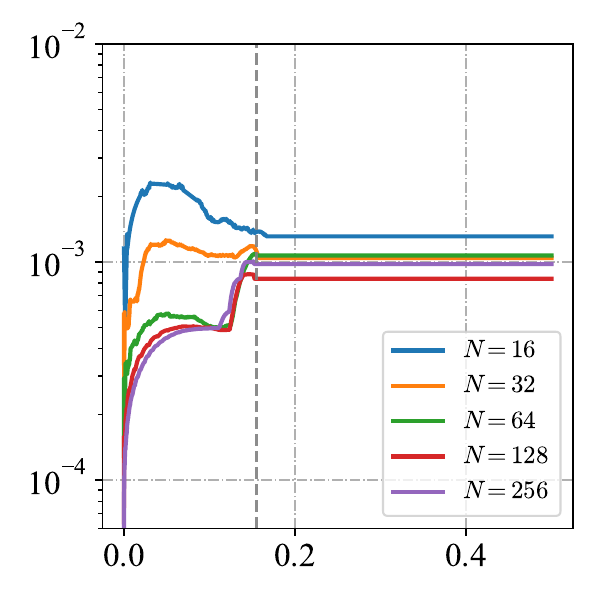}}
        \put(10, 175){(a)}
        \put(85, -2){{time $\axissi{s}$}}
        \put(-5, 55){\rotatebox{90}{ mass error $e_{m}$}}
        \put(77,24){{$t_{\text{s}}$}}

        \put(200, 0){\includegraphics[width=0.33\textwidth]{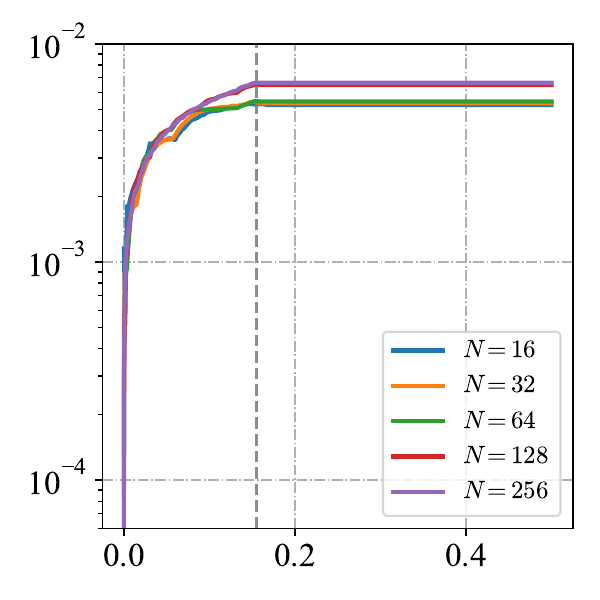}}
        \put(210, 175){(b)}
        \put(285, -2){{time $\axissi{s}$}}
        \put(195, 55){\rotatebox{90}{ mass error $e_{m}$}}
        \put(277,24){{$t_{\text{s}}$}}
    }
    \end{picture}
    \caption{Relative mass error in the horizontal solidification problem for various grid resolutions: (a) without non-conservative bounding  and (b) with non-conservative bounding.}
    \label{fig:hor-case-mesh-conv}
\end{figure}

As shown in Fig.~\ref{fig:hor-case-mesh-conv}, the accumulation of $e_m(t)$ continues until complete solidification, which occurs for the both cases at approximately $t_\text{s}=\SI{0.15}{\s}$. Comparing the results of Figs.~\ref{fig:hor-case-mesh-conv}a,b, non-conservative bounding appears to introduce additional error $e_m$. In the worst-case scenario of mass accumulation presented in Fig.~\ref{fig:hor-case-mesh-conv}b, the mass difference $e_m(t)$ at each time step does not exceed $\num{e-4}$. The unbounded nature of mass accumulation can lead to significant errors in some cases, especially during long simulations.

\begin{figure}[!t]
    \begin{picture}(400, 190)
    \put(5,0){
        \put(0, 0){\includegraphics[width=0.33\textwidth]{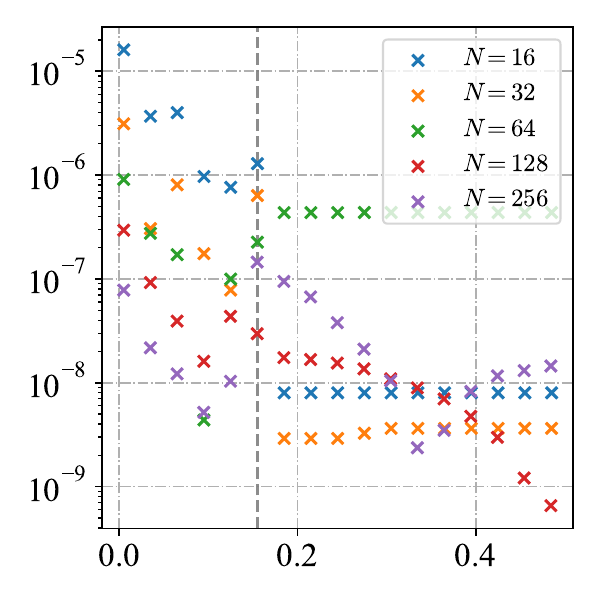}}
        \put(10, 175){(a)}
        \put(85, -2){{time $t$ $\axissi{s}$}}
        \put(-5, 55){\rotatebox{90}{ mass error $e_{m}$}}
        \put(77,24){{$t_{\text{s}}$}}

        \put(200, 0){\includegraphics[width=0.33\textwidth]{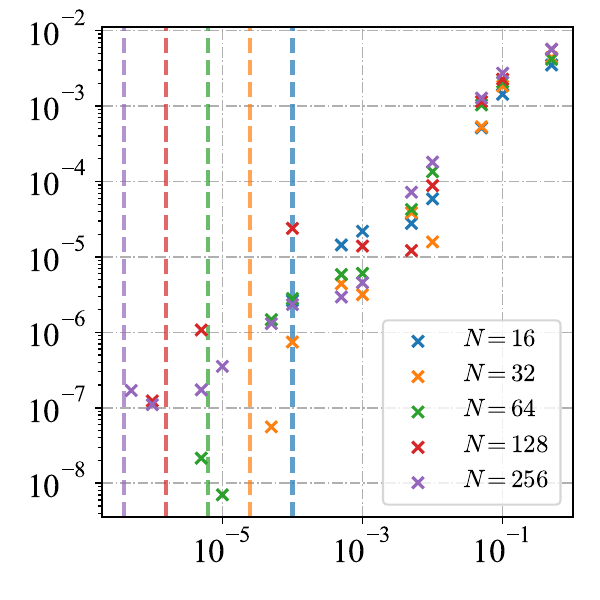}}
        \put(210, 175){(b)}
        \put(260, -2){{correction time $\tau\ \axissi{s}$}}
        \put(195, 55){\rotatebox{90}{ mass error $e_{m}$}}
    }
    \end{picture}
    \caption{\revB{Mass conservation for the horizontal solidification problem for various grid resolutions ($N$): (a) Mass error as a function of time $t$ for  $\tau = \Delta{t}$; (b) The dependence of mass error at $t=0.5$ on the mass-correction parameter $\tau$. The vertical dashed lines correspond to: (a) the complete solidification time $t=t_\text{s}$ and (b)  the time step $\tau=\Delta{t}$.}}
    \label{fig:hor-case-mesh-conv-corr}
\end{figure}
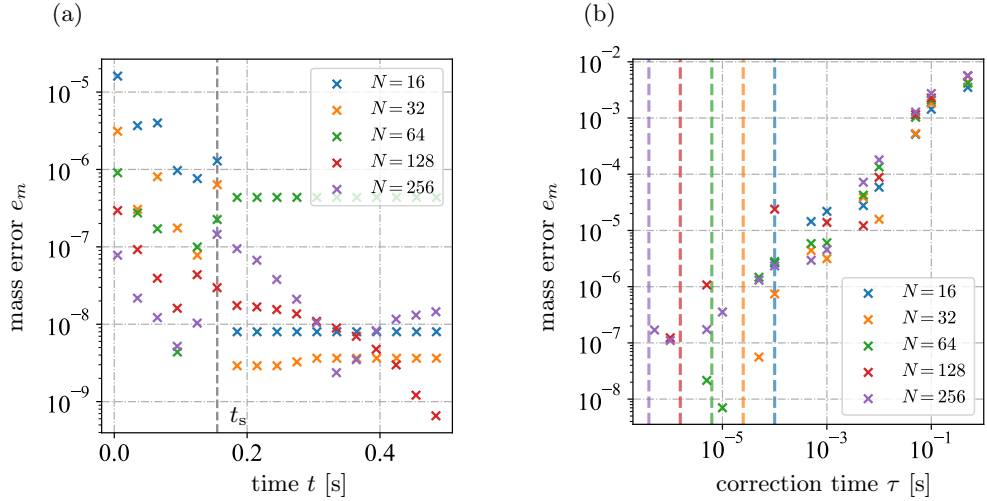

The cumulative mass error is primarily influenced by small stepwise changes, with  each step contributing, on average,  to the overall mass error on the order of $10^{-5}$ and lower. As shown in Fig.~\ref{fig:hor-case-mesh-conv-corr}, the application of mass correction significantly mitigates these errors, leading to improved accuracy. \revB{Decreasing the mass-correction parameter $\tau$ (see Eq.~\eqref{eq:rho_corr}) steadily lowers the mass error until $\tau$ approaches the size of the simulation time step $\Delta{t}$. The use of smaller values of $\tau$ would unnecessarily stiffen the system and could only be used in combination with implicit time integration methods. Thus, for all practical purposes, the parameter $\tau$ can be taken to be the time step.} In simulations with mass correction, the highest mass error $e_m$ is found to be comparable to the mass error over a few time steps, resulting in $e_m$ that is about three to six orders of magnitude lower than without mass correction, as illustrated in Fig.~\ref{fig:hor-case-mesh-conv-corr} (a).

Similar issues with mass conservation have been demonstrated or mentioned in studies where the phase densities were not considered constant \cite[e.g.,][]{Thirumalaisamy2023,Miller2013UDE,Koch2016bublecollapse}. However, these issues largely remained unaddressed, except in the work of Koch et al.~\cite{Koch2016bublecollapse}, who applied a correction by directly manipulating the density of the gas phase. The mass correction procedure,  proposed in Sec.~\ref{sec:numerical}, is particularly beneficial for long simulations on engineering-scale meshes, where mass conservation errors can accumulate over time and significantly affect the accuracy of gas--metal interface position.

\subsection{2D vertical solidification}

As the test problems increase in physical complexity, the influence of capillary effects on the solidification in the presence of a free gas--melt interface is examined next.
The test problem models the formation of shrinkage pipe defects during the solidification of aluminum and is inspired by Ref.~\cite{Thirumalaisamy2023}. This section examines the effects of volume change due to phase transformation and thermal dilatation of the liquid phase on the formation of the metal surface. Additionally, the influence of smoothing of the liquid volume fraction on the location of the free surface and the resulting surface topography is discussed.

\subsubsection{Problem formulation}

Initially, a rectangular domain $\Omega \in [0, l]\cross [0, l]$, where  $l = \SI{8e-4}{\m}$, is filled with liquid aluminum at a temperature twice its melting point, up to a height of \SI{5e-4}{\m}. The initial conditions for temperature and velocity  mirror those used in the previous problem.  The side boundaries of the domain are subject to  isothermal and no slip velocity boundary conditions. Adiabatic and stress-free boundary conditions are applied to the bottom boundary of the domain, while the top boundary  has open conditions for both velocity and pressure, with the temperature set to $T=0.5T_\melt$, as illustrated in Fig.~\ref{fig:comparison-of-density-jump}. The boundary conditions are summarized as follows:
\begin{align}
    \begin{split}
        {k\pdv{T}{y}}(x,0,t) &= q_0,\quad  T(x,l_y,t) = T(0,y,t) = T(l_x,y,t) = 0.5T_\melt,\\
        \bv(x,0,t) &= \bv(0,y,t) = \bv(l_x,y,t) = \bv_0,\quad p(x,l_y,t) = p_0,
    \end{split}
\end{align}
where $q_0 = \SI{0}{\W\per\m^2}$,   $|\bv_0| = \SI{0}{\m\per\s}$, and  $p_0 = \SI{0}{\Pa}$.

\subsubsection{Results and discussion}

The effects of density variation due to phase transformation as well as thermal dilatation are demonstrated by comparing the following three scenarios:
\begin{enumerate}[topsep=10pt, partopsep=0pt,itemsep=3.0pt,parsep=2pt,leftmargin=30pt,labelsep=10pt, label = (\arabic*)]
    \item constant metal density, $\rho_\liq = \rho_{\liq\melt} = \rho_\sol$,
    \item density jump due to phase transition, $ \rho_\liq = \rho_{\liq\melt}\neq \rho_\sol$,
    \item variable density, $\rho_\liq = \rho_{\liq\melt} + \rho_{\liq\melt}\beta(T-T_\melt)$, $\rho_\sol \neq \rho_{\liq\melt}$.
\end{enumerate}
The results are obtained using a mesh resolution of $N \times N$, where $N=256$, and a time step of $\Delta t = \SI{e-6} {\s}$. In each scenario, mass correction is applied. The comparison of results for the three scenarios of density change is  presented with piecewise liquid volume fraction. Similar to the previous case, the relative mass error does not exceed $e_m \leq \num{e-5}$. The value of the penalization parameter $\epsilon$ is set to $\num{e-7}$. The gravitational source term in Eq.~\eqref{eq:momentum_conservation} is neglected.

\begin{figure}[t]
    \centering
    \begin{picture}(400, 320)
    \put(18, 0){
        \put(0, 10){\includegraphics[width=0.725\linewidth]{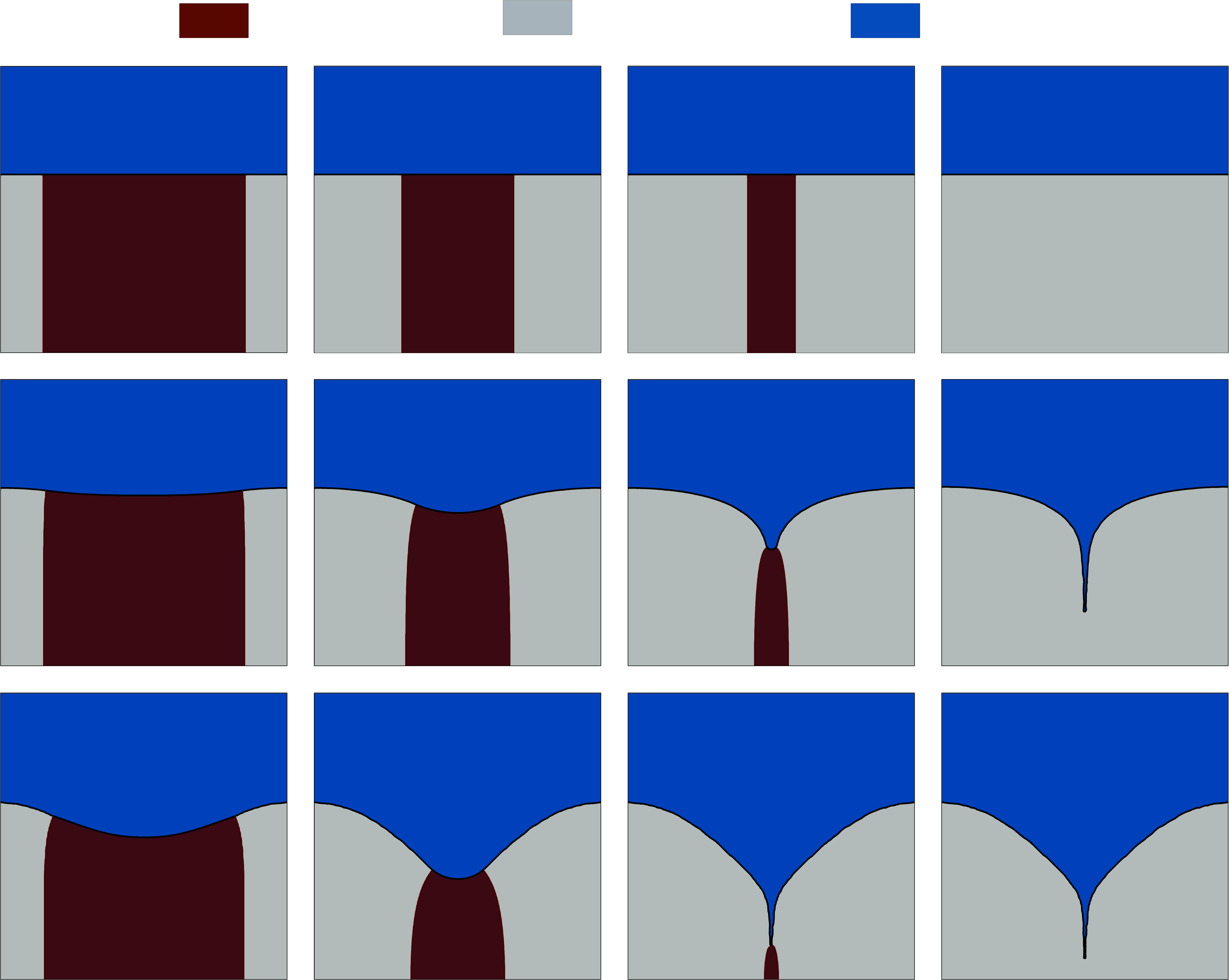}}
        \put(85, 297){
         -- \quad 
        \begin{tabular}{@{}c@{}} 
            molten \\ 
            metal     
        \end{tabular}%
    }
        \put(185, 297){ -- \quad
        \begin{tabular}{@{}c@{}} 
            solidified  \\ 
            metal     
        \end{tabular}%
        }
        \put(285, 297){ -- \quad gas}

        \put(45, 1){\makebox(0,0)[c]{ $t = 0.05\;\pictsi{s}$}}
        \put(143, 1){\makebox(0,0)[c]{ $t = 0.15\;\pictsi{s}$}}
        \put(235, 1){\makebox(0,0)[c]{ $t = 0.20\;\pictsi{s}$}}
        \put(331, 1){\makebox(0,0)[c]{ $t = 0.5\;\pictsi{s}$}}

        \put(-18, 276){(a)}
        \put(-18, 182){(b)}
        \put(-18, 87){(c)}
     }
    \end{picture}
    \caption{Vertical solidification problem for aluminum and its numerical solution, illustrating the phase distribution at different times for different density variation scenarios: (a) equal constant densities $\rho_\liq = \rho_\sol$; (b) different constant densities $\rho_\liq \ne \rho_\sol$; (c) temperature-dependent liquid metal density $\rho_\liq(T)$ and constant solid metal density $\rho_\sol$.}
    \label{fig:comparison-of-density-jump}
\end{figure}
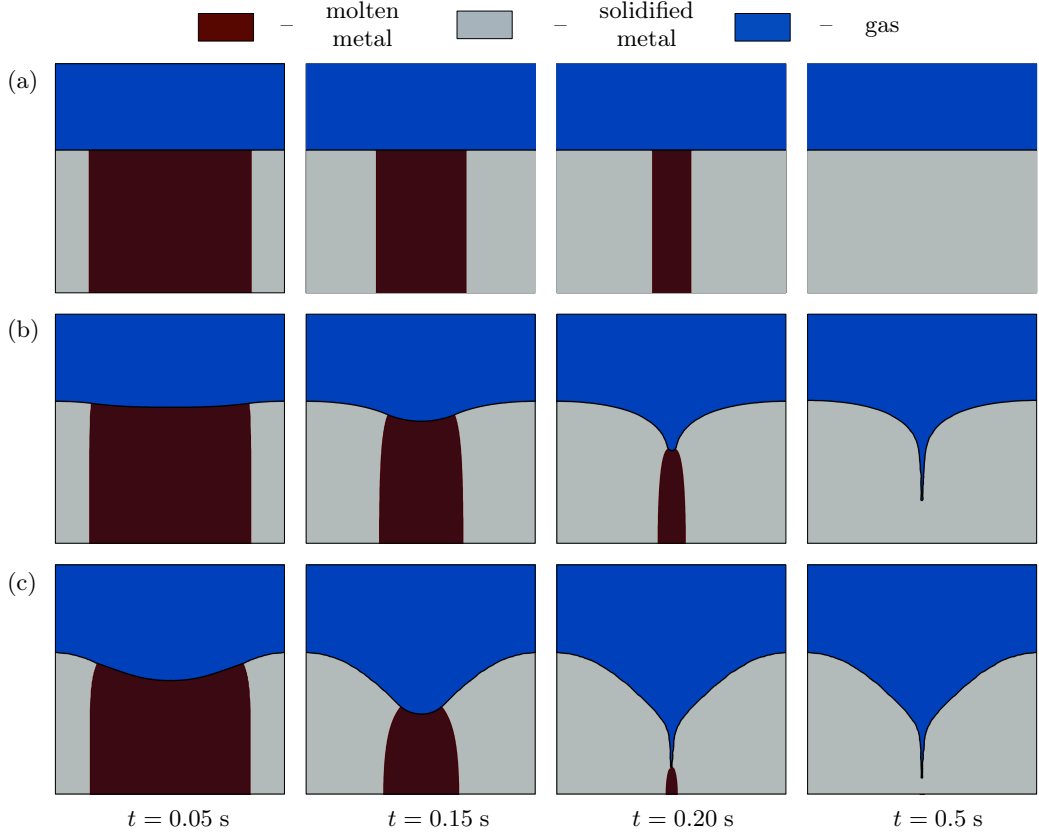

As shown in Fig.~\ref{fig:comparison-of-density-jump}, there is no movement of the gas--metal interface in the first scenario. However, in the second and third scenarios, the gas--metal interface changes significantly. Both of these cases exhibit a narrow gap at the center of the domain, commonly referred to as a solidification crack or solidification pipe. Additionally, a difference in interface depression due to thermal dilatation is observed in the second and third scenarios, which highlights the importance of considering variable metal density and its temperature dependence for accurate prediction of the final surface topography.

Upon examining the solution to this problem, several observations can be made. In the second and third scenarios, an interface configuration develops where the liquid fraction becomes completely surrounded by the solid fraction and the boundary. As the crack narrows, its geometry cannot be accurately resolved with the given mesh resolution, resulting in the liquid metal being enclosed by solid metal. As discussed in Sec.~\ref{sec:assumptions}, the occurrence of liquid being entirely trapped within the solid violates basic assumptions of the model, resulting in non-physical behavior. However, since the penalization coefficient remains finite in the solid region, it  allows for slight movement in the solid phase, thus enabling the pressure equation to be solved. By relaxing the penalization coefficient to $\epsilon = \num{e-7}$, crack formation can be maintained, and the simulation can continue even when the liquid is enclosed by the solid.

A critical aspect of this problem is the emergence of parasitic current near the gas--melt interface,  which results from improper estimation of surface curvature and can significantly affect the position of the surface, especially on a fine mesh. The most widely used method for curvature estimation method, which relies on the gradient of the metal volume fraction $\alpha$, has been shown not to converge with mesh refinement in the implementation of geometric VOF~\cite{Scheufler2023Roenby}.  To mitigate this issue,  two approaches have been utilized. The first involves omitting the modified gradient $\gradf{f}$ for the surface tension term in Eq.~\eqref{eq:v_scheme}, which results in a drastic reduction of effect of spurious velocities on the gas--melt interface. The second approach is application of reconstructed distance function for surface curvature calculation methods, as discussed by Roenby et al.~\cite{Scheufler2023Roenby}.

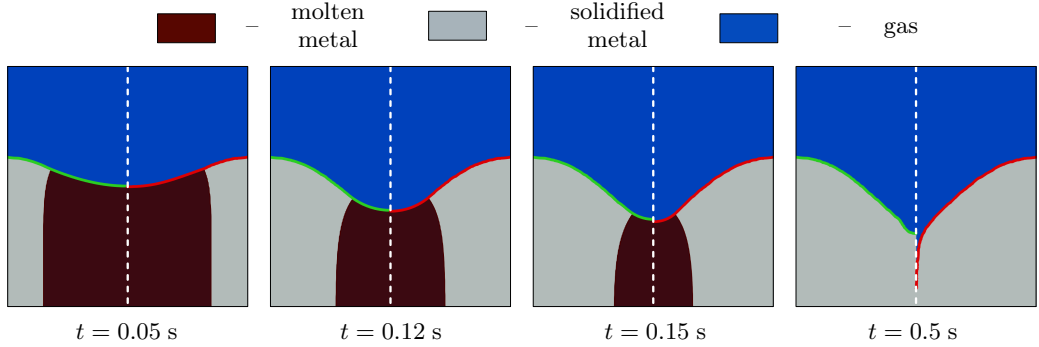
\begin{figure}[t]
    \centering

    \begin{picture}(400, 130)
        \put(0, 10){\includegraphics[width=0.76\linewidth]{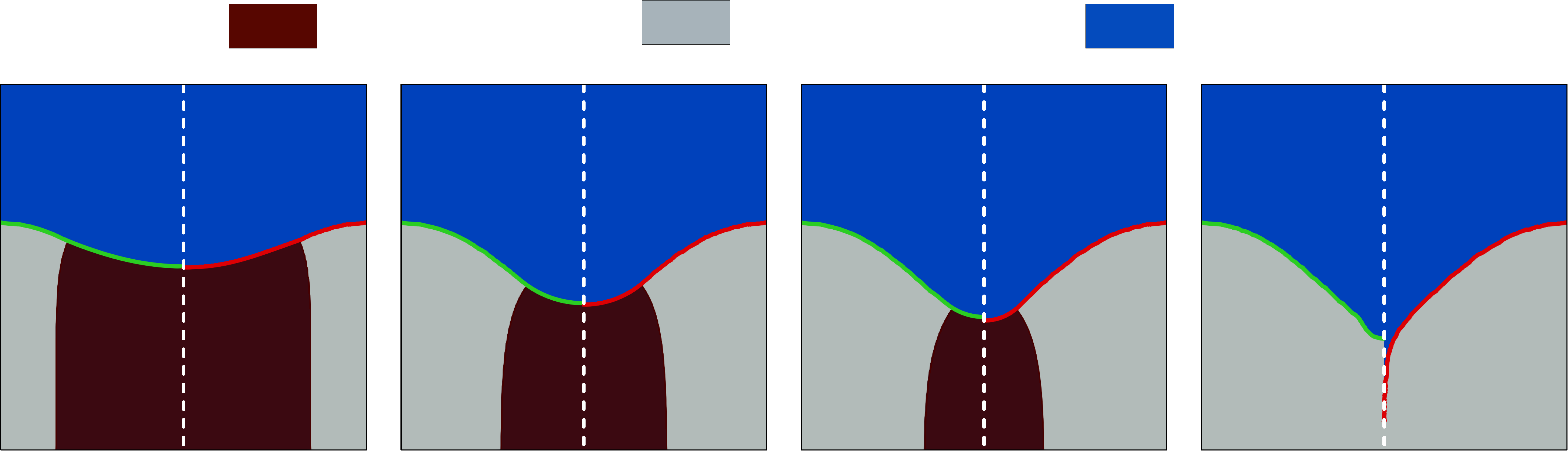}}
        \put(90, 113){
         -- \quad 
        \begin{tabular}{@{}c@{}} 
            molten \\ 
            metal     
        \end{tabular}%
    }
        \put(192, 113){ -- \quad
        \begin{tabular}{@{}c@{}} 
            solidified  \\ 
            metal     
        \end{tabular}%
        }
        \put(310, 113){ -- \quad gas}

        \put(43.5, 1){\makebox(0,0)[c]{ $t = 0.05\;\pictsi{s}$}}
        \put(143, 1){\makebox(0,0)[c]{ $t = 0.12\;\pictsi{s}$}}
        \put(242.5, 1){\makebox(0,0)[c]{ $t = 0.15\;\pictsi{s}$}}
        \put(339.5, 1){\makebox(0,0)[c]{ $t = 0.5\;\pictsi{s}$}}
    \end{picture}
    \caption{\revC{Comparison of the interface in the vertical solidification problem evaluating $\phi_\met$ using Eq.~\eqref{eq:phi} (green line) and  Eq.~\eqref{eq:erf_sigmoid} (red line).}}
    \label{fig:cut-erf-interface-comparison}
\end{figure}

The application of a smooth sigmoid-based function for evaluating the liquid volume fraction form the enthalpy also affects the surface formation compared to  using a piecewise function. In the following example, the result of the vertical solidification problem are compared for piecewise and sigmoid-based functions, defined by Eq.~\eqref{eq:phi} and Eq.~\eqref{eq:erf_sigmoid}, respectively. The effect of locally smoothing the liquid volume fraction is demonstrated by comparing the phase distribution during the solidification process for these two functions. As shown in Fig.~\ref{fig:cut-erf-interface-comparison}, the difference in phase distribution is minor until the final stage of solidification, when the solidification pipe forms.

\subsection{Melting in a rectangular container}

The following validation problem addressed melting in a rectangular container and is based on the classic experimental study by Gau and Viskanta~\cite{Gau1986Viskanta} along with computational studies by Brent et al.~\cite{Brent1988EnthalyPorosity} and Hannoun et al.~\cite{Hannoun2003}. The problem accounts for convective flow induced by temperature-dependent density variations. Unlike the Boussinesq approximation, in this test case, convection is modeled through an explicit density-temperature relationship.

\subsubsection{Problem formulation}
The melting process is considered in a two-dimensional rectangular domain $\Omega \in [0, l_x] \times [0, l_y]$, where $l_x = \SI{88.9}{\mm}$ and $l_y = \SI{79.375}{\mm}$. Initially, the container is partially filled with solid metal up to height  $h = \SI{63.5}{\mm}$ at temperature $T_0 = \SI{301.3}{\K}$, with the remaining space occupied by gas  at the same temperature. Melting initiates from the wall maintained at $T_1 = \SI{311.0}{\K}$. The boundary conditions are specified as follows:
\begin{align}
    \begin{split}
        T(0,y,t) &= T_1, \quad T(l_x,y,t) = T_0,\quad {k\pdv{T}{y}}(x,0,t) = {k\pdv{T}{y}}(x,l_y,t) = q_0,\\
        \bv(x,0,t) &= \bv(0,y,t) = \bv(l_x,y,t) = \bv_0,\quad p(x,l_y,t) = p_0,
    \end{split}
\end{align}
where $q_0 = \SI{0}{\W\per\m^2}$, $|\bv_0| = \SI{0}{\m\per\s}$, and $p_0 = \SI{0}{\Pa}$. The thermophysical properties for the metal match those used by Hannoun et al.~\cite{Hannoun2003} and parameters for gas are summarized in Table~\ref{Tab:al_thermophysical}.  The gas viscosity  is  increased by three orders of magnitude to enable  the gas phase to act as an adiabatic no-slip wall while allowing for the thermal expansion of the metal.

\subsubsection{Results}
\begin{figure}[!t]
    \centering
    \begin{picture}(200, 180)
    \put(0, 0){\includegraphics[width=0.41\textwidth]{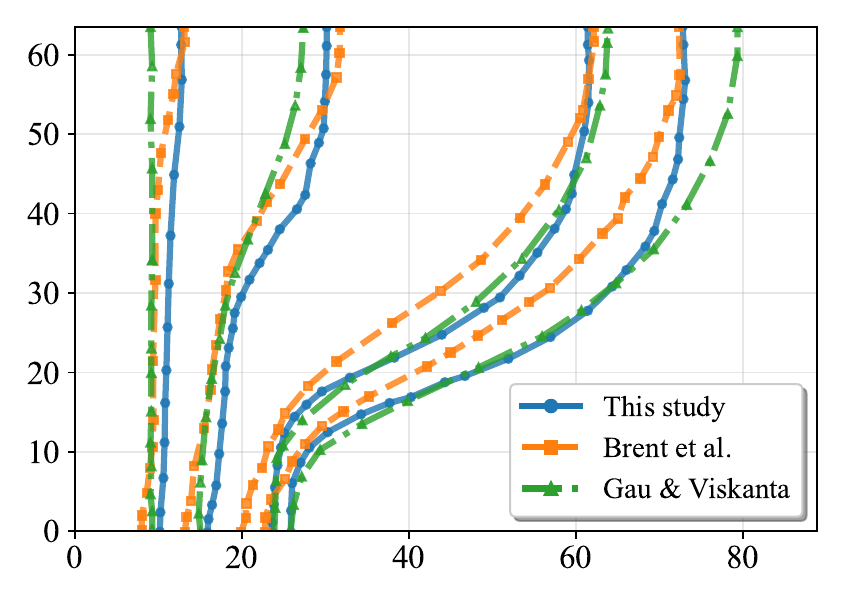}}
        \put(90, -3){{width $\axissi{\mm}$}}
        \put(-7, 55){\rotatebox{90}{ height $\axissi{\mm}$}}
        \put(30,150){{2 min}}
        \put(70,150){{6 min}}
        \put(132,150){{15 min}}
        \put(168,150){{19 min}}
    \end{picture}
    \caption{Comparison of melt-solid interface positions at various times with  reference results from Refs.~\cite{Gau1986Viskanta,Brent1988EnthalyPorosity}.}
    \label{fig:ga-coarse-mesh}
\end{figure}
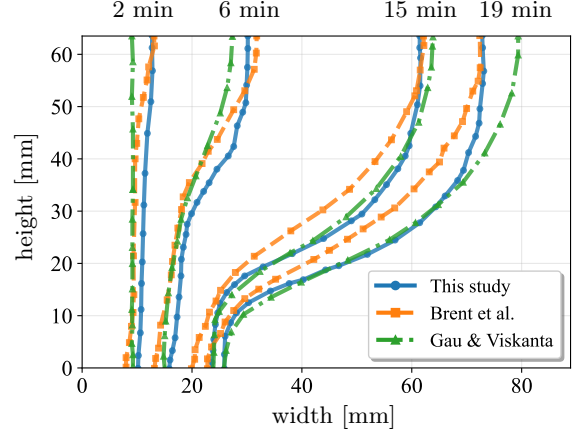

The gallium melting problem and results from Brent et al.~\cite{Brent1988EnthalyPorosity} are widely regarded as reliable benchmarks for validating numerical simulations~\cite{Kozak2017,Kewalramani2020,roache1998verification}, although some concerns have been raised about the experimental study and numerical simulations described by Hannoun et al.~\cite{Hannoun2003}. A comparison of coarse grid simulations between the present framework with those from  Gau and Viskanta~\cite{Gau1986Viskanta} and Brent et al.~\cite{Brent1988EnthalyPorosity}, as shown in Fig.~\ref{fig:ga-coarse-mesh}, demonstrates good agreement.

\begin{figure}[!t]
    \centering
    \begin{picture}(200, 250)
    \put(0, 52){\includegraphics[width=0.41\textwidth]{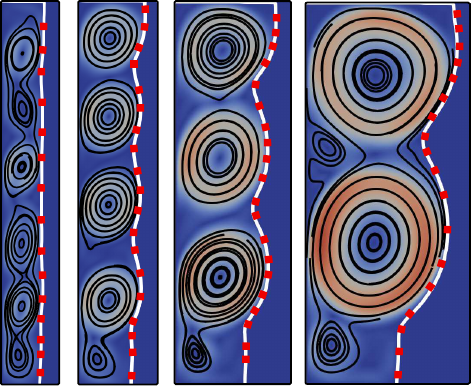}}
        \put(14,236.5){{{\makebox(0,0)[c]{\includegraphics[width=0.06\textwidth]{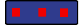}}}}}
        \put(69,236.5){\makebox(0,0)[c]{{-- Hannoun et al.}}}
        \put(127,236.5){{{\makebox(0,0)[c]{\includegraphics[width=0.06\textwidth]{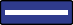}}}}}
        \put(180,236.5){\makebox(0,0)[c]{{-- present study}}}
        \put(16,46){\makebox(0,0)[c]{\scriptsize{42 s}}}
        \put(56,46){\makebox(0,0)[c]{\scriptsize{85 s}}}
         \put(109,46){\makebox(0,0)[c]{\scriptsize{155 s}}}
          \put(178,46){\makebox(0,0)[c]{\scriptsize{280 s}}}
        \put(110, 28){{\makebox(0,0)[c]{\includegraphics[width=0.30\textwidth]{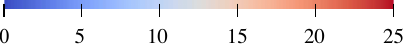}}}}
        \put(110,7){{{\makebox(0,0)[c]{{velocity magnitude [$\si{\mm\per\s}$]}}}}}
    \end{picture}
    \caption{Streamlines and melt--solid interface at various times. The latter is compared with results from Ref.~\cite{Hannoun2003}.}
    \label{fig:ga-fine-mesh}
\end{figure}
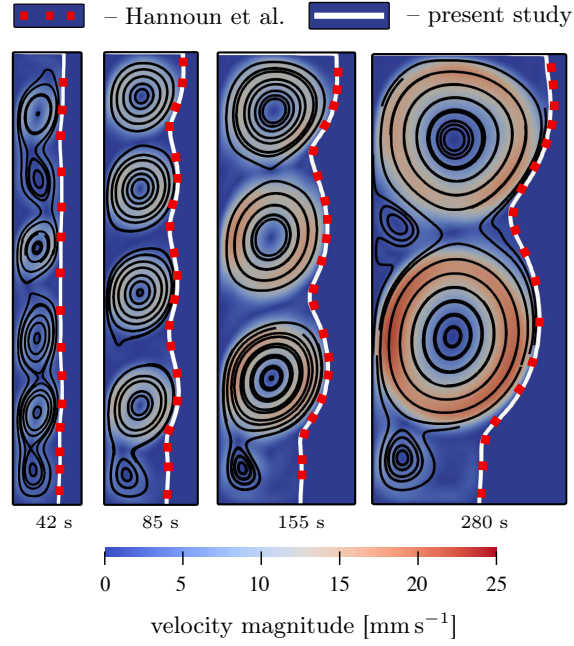

A grid-converged solution from Hannoun et al.~\cite{Hannoun2003}  is used as the reference solution for comparing melt dynamics and phase front position.  The results of the present study, with  a grid resolution of $\num{560} \times \num{400}$ in the molten metal region, show excellent agreement with the reference solution, as seen in Fig.~\ref{fig:ga-fine-mesh}. These results confirm that the proposed model accurately reproduces the position of melting interfaces in the presence of free convection.

\subsection{Pulsed laser melting}
The final test problem demonstrates a two-phase metal solidification process in the presence of the capillarity, Marangoni force, and thermal dilatation, while considering surface energy deposition from an incident laser beam modeled as a surface heat source. This demonstration case is based on the study by Pitscheneder et al.~\cite{Pitscheneder1996}, which has also been used as a test case for developing numerical simulation models for metal melting  due to an incident laser beam in Refs.~\cite{saldi2012marangoni, Ebrahimi2020, Yan2018}. The influence of compressibility on the location of the gas--metal interface during and after melting is highlighted.

\begin{figure}[!t]
    \begin{picture}(400, 260)
    \put(5,0){
        \put(0, 40){\includegraphics[width=0.34\textwidth]{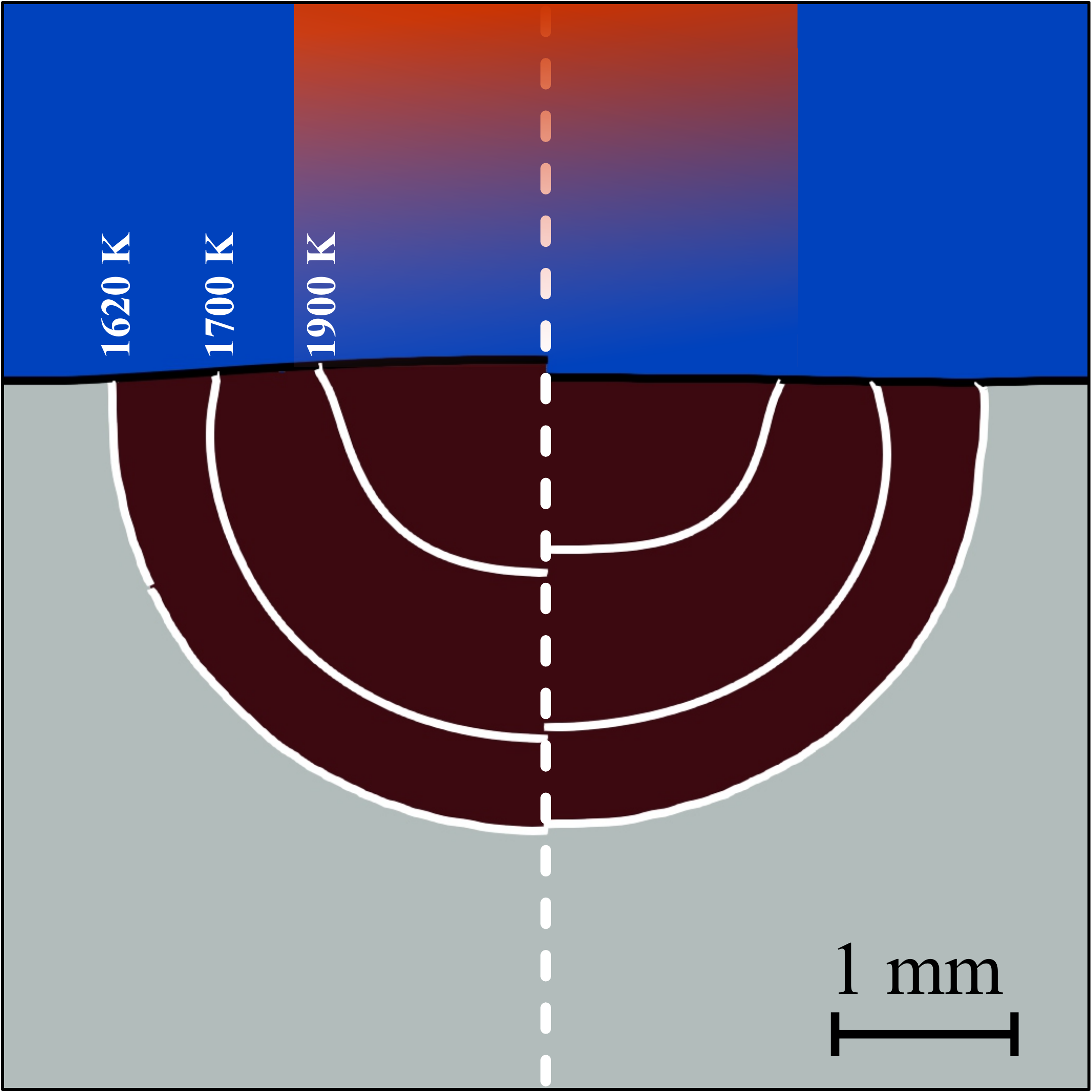}}
        \put(15, 220){$\rho_\liq (T) \neq \rho_\sol$}
        \put(105, 220){$\rho_\liq = \rho_\sol$}
        \put(89, 243){\makebox(0,0)[c]{150 ppm}}

        \put(200, 40){\includegraphics[width=0.34\textwidth]{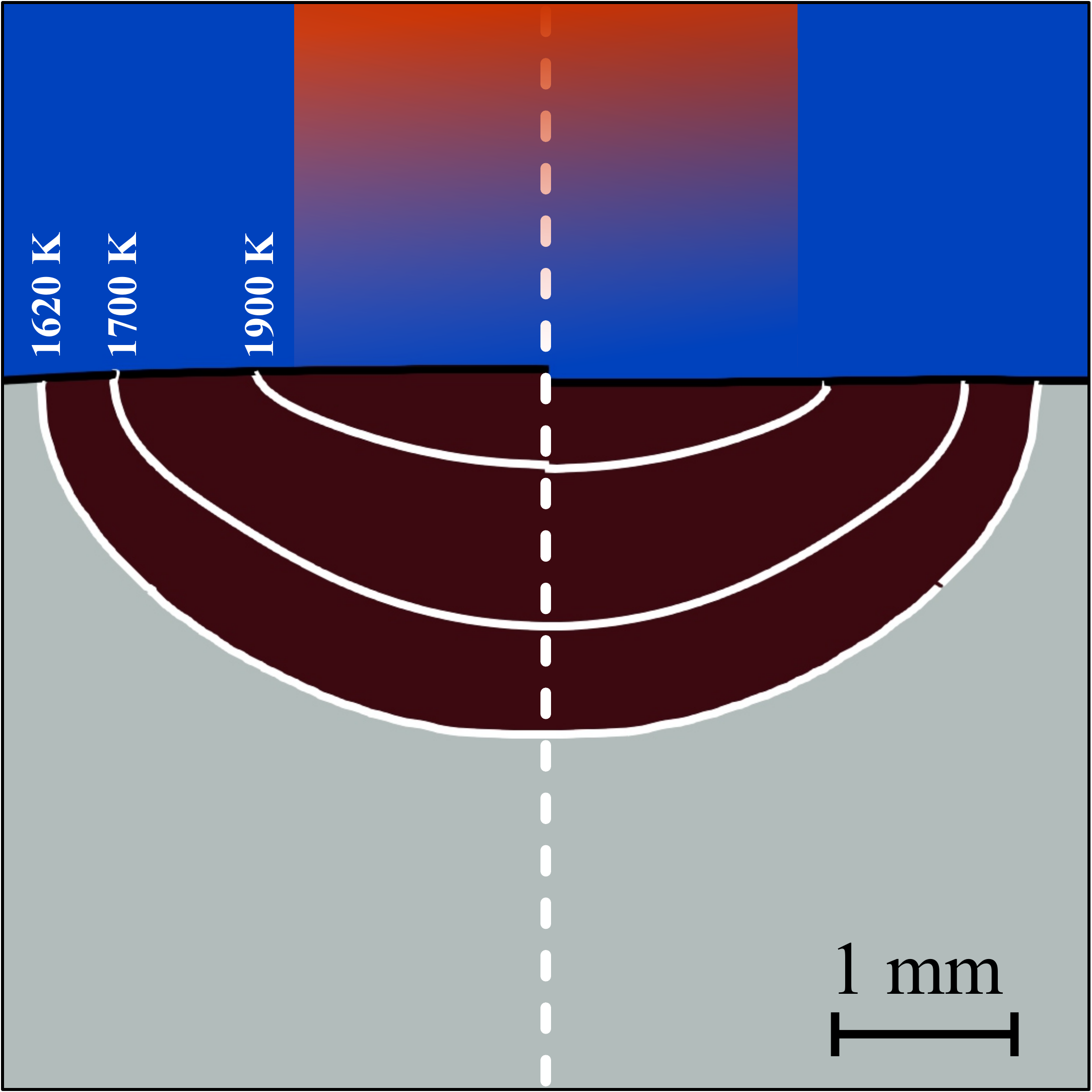}}
        \put(215, 220){$\rho_\liq (T) \neq \rho_\sol$}
        \put(305, 220){$\rho_\liq = \rho_\sol$}
        \put(289, 243){\makebox(0,0)[c]{20 ppm}}

        \put(55, 9){\includegraphics[width=0.42\linewidth]{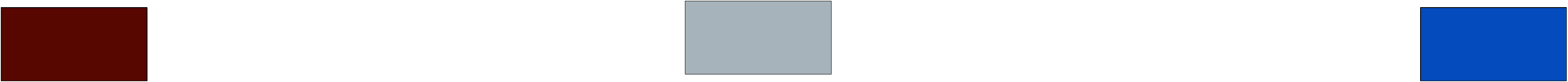}}
        \put(85, 12){
         -- \quad 
        \begin{tabular}{@{}c@{}} 
            molten \\ 
            metal     
        \end{tabular}%
    }
        \put(185, 12){ -- \quad
        \begin{tabular}{@{}c@{}} 
            solidified  \\ 
            metal     
        \end{tabular}%
        }
        \put(285, 12){ -- \quad gas}
    }
    \end{picture}
    \caption{Comparison of numerically simulated phase distribution and temperature contours at the cross-section after 5 seconds for variable $\rho_\liq (T) \neq \rho_\sol$ and constant $\rho_\liq = \rho_\sol$ liquid metal densities for different sulfur concentrations.}
    \label{fig:Pitscheneder_sim_comparison}
\end{figure}
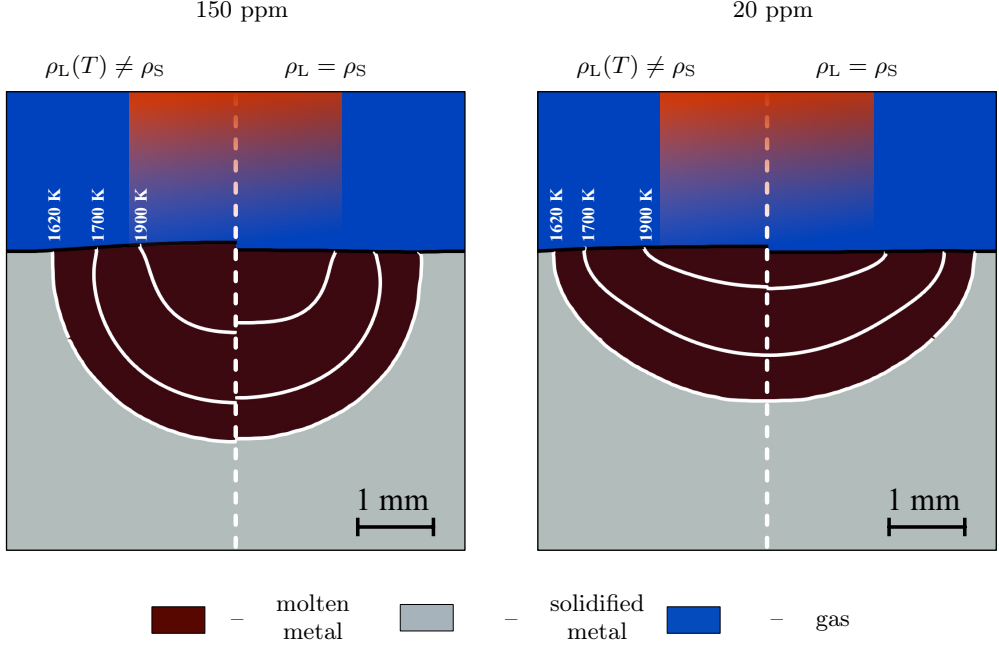

\subsubsection{Problem formulation}

The pulsed laser melting process is modeled as an axisymmetric problem and  solved in a cylindrical domain with a radius of $r_0 = \SI{8}{\mm}$ and a height of $h = \SI{16}{\mm}$.  The solution is assumed to be independent of the  azimuthal direction.
Initially, the bottom half of the domain is filled with metal at room temperature, $T_0 = \SI{293}{K}$, with both the gas and the metal being at rest.  Stress-free and open boundary  conditions are applied at the top of the domain.  Axisymmetric boundary conditions are applied at the axis of symmetry.
The other boundary conditions are formulated as follows:
\begin{align}
    \begin{split}
        T(r,z=h) &= T(r,z=0) = T (r=r_0,z) = T_0,    \\
        p(r,z=h) &= p_0,\quad \bv(r,z=0) = \bv (r=r_0,z) = \bv_0,
    \end{split}
\end{align}
where $|\bv_0| = \SI{0}{\m\per\s}$ and  $p_0 = \SI{e5}{\Pa}$.
The energy from an incident laser beam is deposited for  5 seconds, with the simulation continuing for a total of 6 seconds to capture the solidification stage. Thermophysical parameters are taken from the original work by Pitscheneder et al. \cite{Pitscheneder1996}.  The liquid density is assumed to vary linearly with temperature as $\rho_\liq = \rho_{\liq\melt}(1 + \beta(T - T_\melt))$, where $\rho_{\liq\melt} = \SI{7900}{\kg\per\cubic\m}$ and $\beta = \SI{-1.01e-04}{\per\K}$. The gravitational source term in Eq.~\eqref{eq:momentum_conservation} is neglected because the Bond number is small, with  $\text{Bo} \approx \num{e-3}$. The calculations are carried out using the sigmoid-based function to evaluate the liquid volume fraction, along with the mass correction procedure. \revB{The cell size in the melting is chosen $\Delta h = \SI{50}{\micro\m}$ with overall number of cells around $\num{2e4}$. The computational time was around 5 hours using 4 cores of Xeon Gold 6326.}

\subsubsection{Results}

In the original study by Pitscheneder et al.~\cite{Pitscheneder1996}, two cases of surface tension  dependence on temperature are considered. The variation in surface tension is attributed to differences in the chemical composition of the steel samples. The dependence of surface tension on chemical composition results in distinct dynamics within the melt pool, influencing phase distribution and the location of the gas--metal interface during melting.

Two density scenarios are compared in this problem: one, where  the solid and liquid metal have equal and constant densities ($\rho_\liq = \rho_\sol$), and another, where the differences due to phase transformation and thermal dilatation are accounted for ($\rho_\liq = \rho_\liq(T)$). In the scenarios considered, which involve melting from an incident laser beam, the  overall melt-pool dynamics is most significantly influenced by the surface tension and its gradient. As a result, only minor differences in the phase distributions are observed in Fig.~\ref{fig:Pitscheneder_sim_comparison}, where the cases with constant and variable densities are compared.

\begin{figure}[!t]
    \centering
    \begin{picture}(200, 190)
    \put(0, 0){\includegraphics[width=0.33\textwidth]{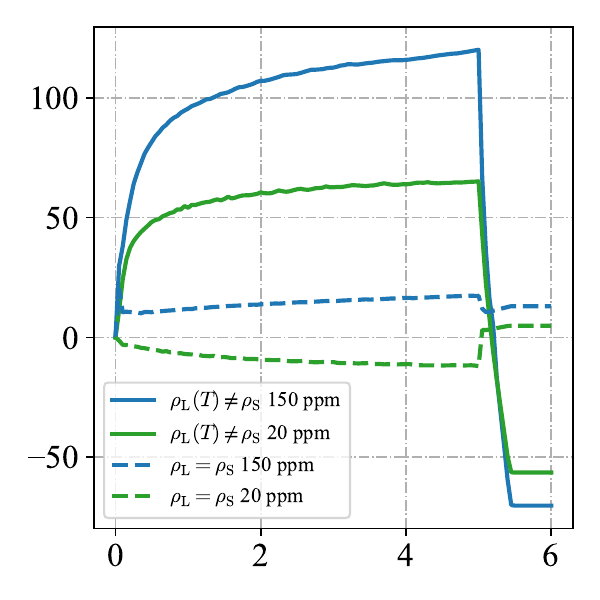}}
        \put(85, -1){{time $\axissi{s}$}}
        \put(-5, 55){\rotatebox{90}{ height $h\; \axissi{\micro \m}$}}
    \end{picture}
    \caption{Comparison of the time evolution of metal surface height along the axis of symmetry  for variable ($\rho_\liq (T) \neq \rho_\sol$) and constant ($\rho_\liq = \rho_\sol$) liquid metal densities for different sulfur concentrations.}
    \label{fig:pitscheneder-height}

\end{figure}
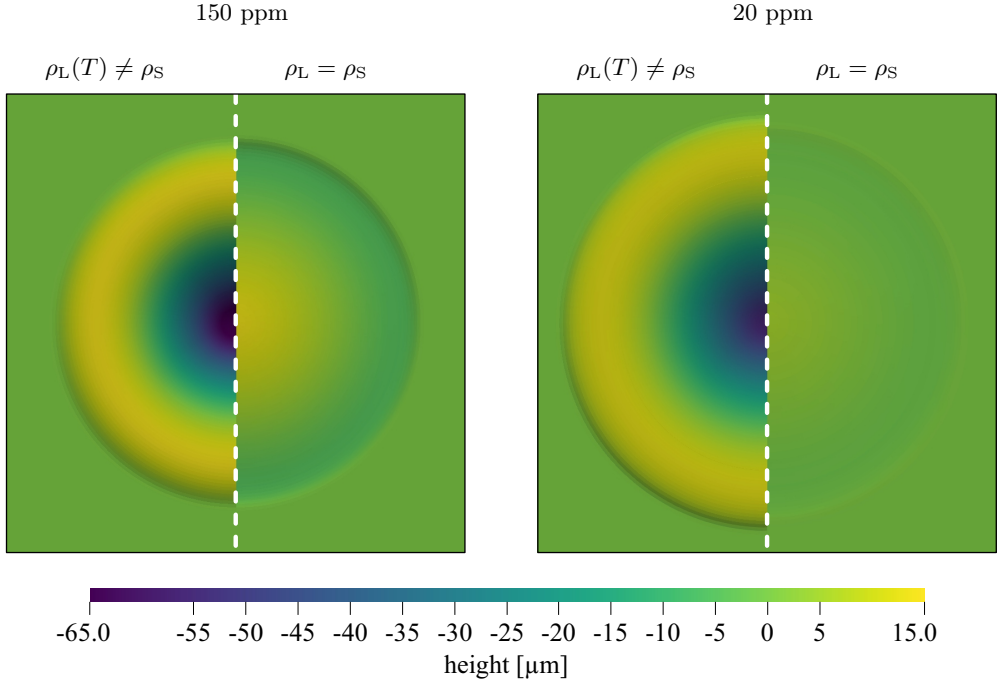
\begin{figure}[!t]
    \begin{picture}(400, 270)
        \put(5,0){
        \put(0, 50){\includegraphics[width=0.34\textwidth]{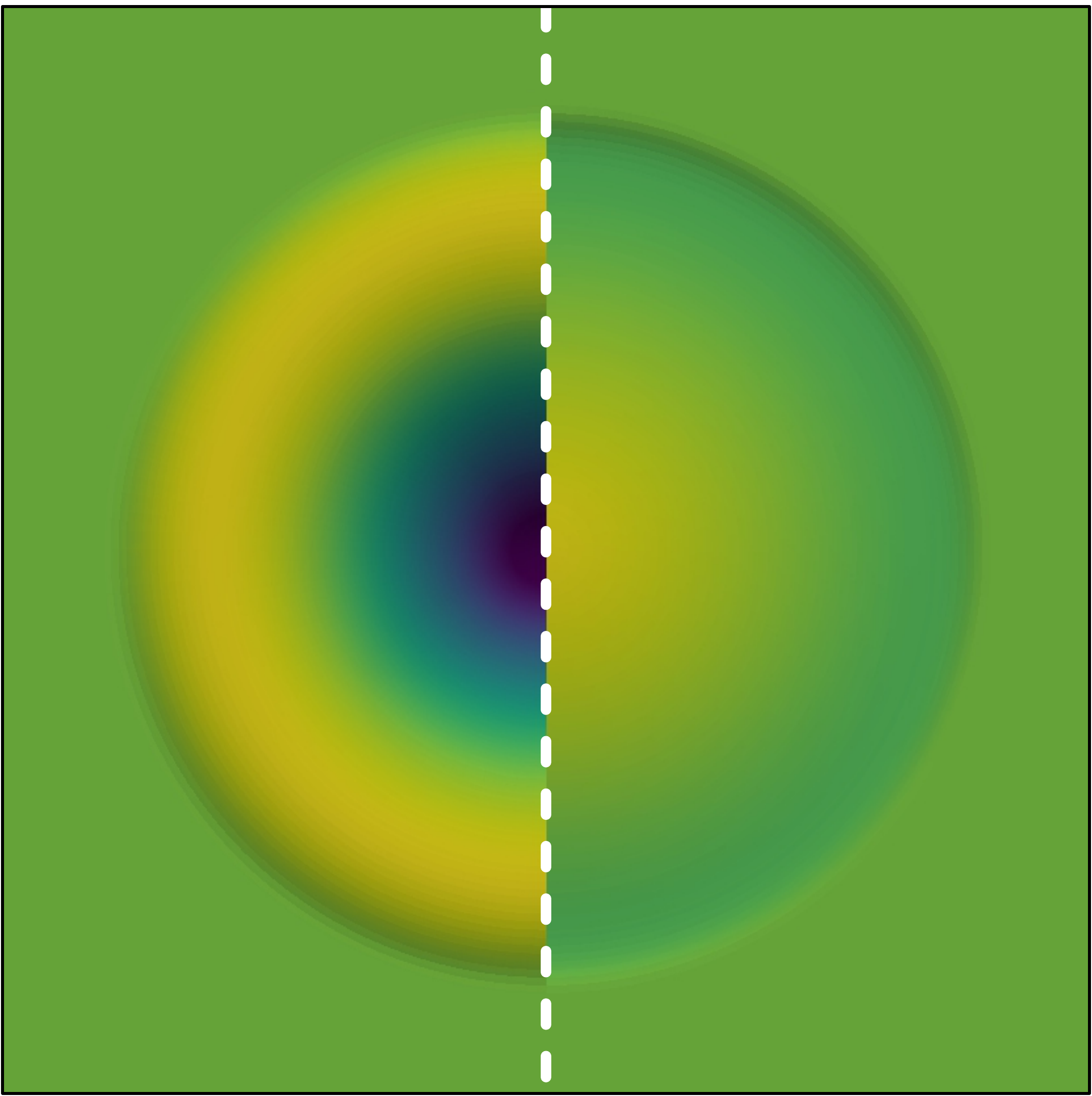}}
        \put(15, 230){$\rho_\liq (T) \neq \rho_\sol$}
        \put(105, 230){$\rho_\liq = \rho_\sol$}
        \put(89, 253){\makebox(0,0)[c]{150 ppm}}

        \put(200, 50){\includegraphics[width=0.34\textwidth]{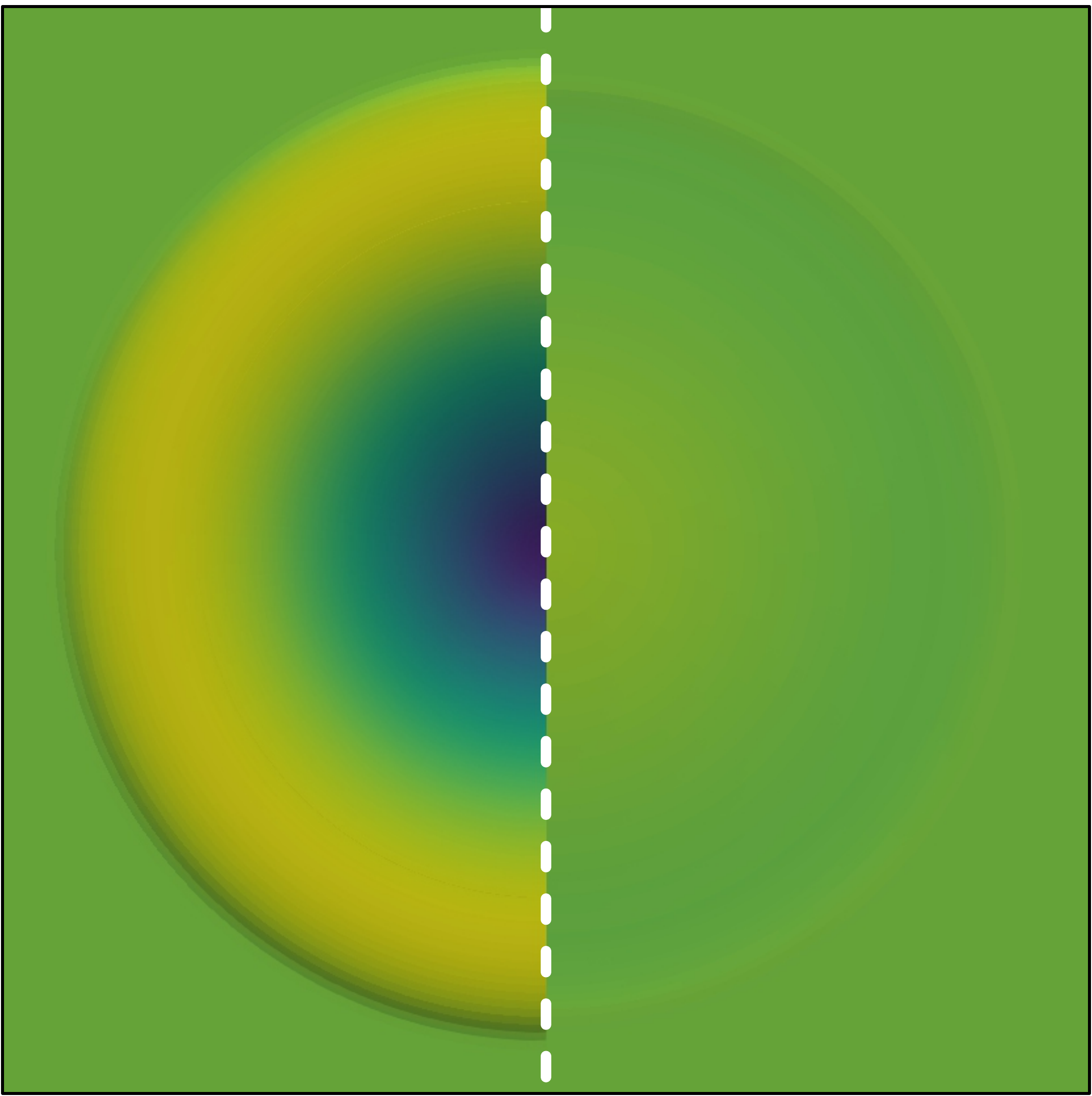}}
        \put(215, 230){$\rho_\liq (T) \neq \rho_\sol$}
        \put(305, 230){$\rho_\liq = \rho_\sol$}
        \put(289, 253){\makebox(0,0)[c]{20 ppm}}

        \put(185, 20){{\makebox(0,0)[c]{\includegraphics[width=0.65\textwidth]{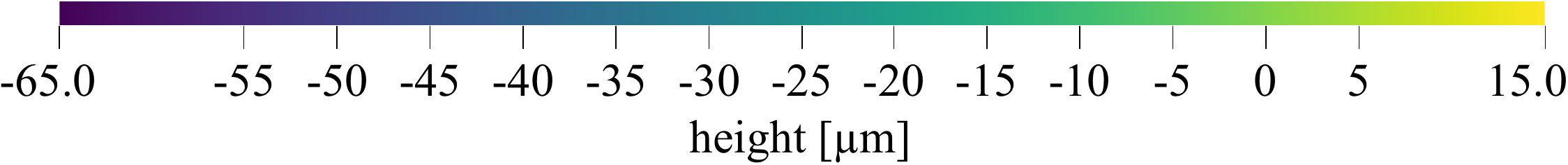}}}}
        }
    \end{picture}
    \caption{Comparison of numerically simulated metal surface topographies after complete solidification for variable ($\rho_\liq(T) \neq \rho_\sol$) and constant ($\rho_\liq = \rho_\sol$) liquid metal densities at different sulfur concentrations.}
    \label{fig:Pitscheneder_melting_topo}
\end{figure}

\begin{figure}[!t]
    \begin{picture}(400, 190)
        \put(5,0){
        \put(80, 40){\includegraphics[width=0.33\textwidth]{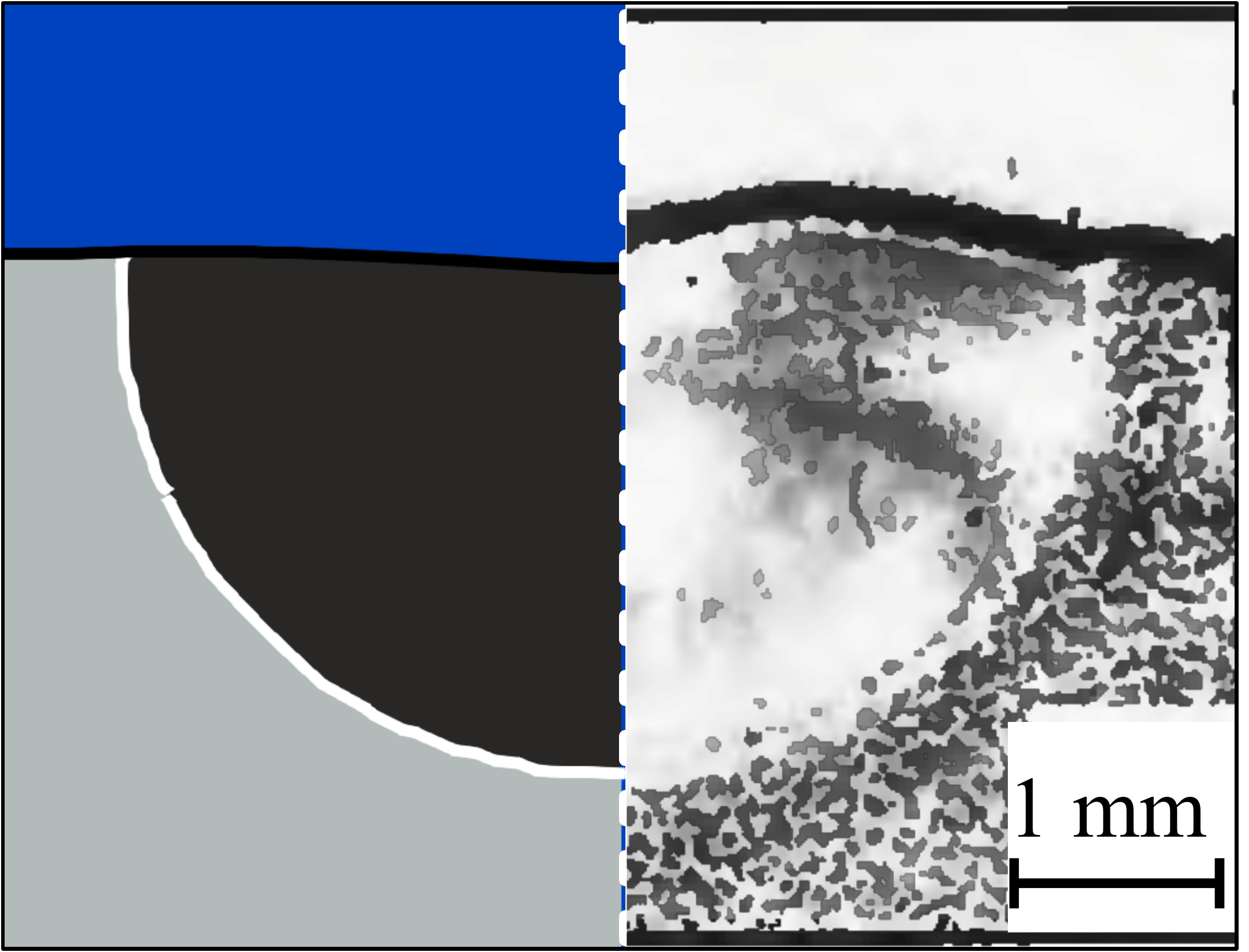}}
                \put(55, 9){\includegraphics[width=0.42\linewidth]{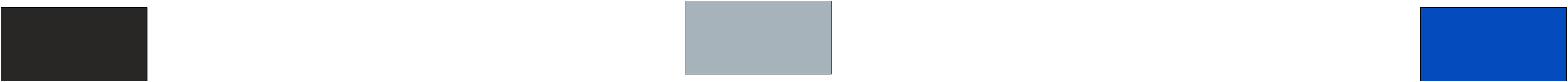}}
        \put(85, 12){
         -- \quad 
        \begin{tabular}{@{}c@{}} 
            resolidified \\ 
            metal     
        \end{tabular}%
    }
        \put(185, 12){ -- \quad
        \begin{tabular}{@{}c@{}} 
            solidified  \\ 
            metal     
        \end{tabular}%
        }
        \put(285, 12){ -- \quad gas}
    }
    \end{picture}
    \caption{\revALL{Comparison of the numerically simulated and experimentally measured~\cite{Pitscheneder1996} phase distributions at the cross-section for 150 ppm sulfur content, following 5 seconds of laser irradiation and subsequent complete solidification.}}
    \label{fig:Pitscheneder_150_pools_comparison}
\end{figure}

The location of the free surface during processing, as well as the final surface topography represented by the gas--metal interface after complete solidification, is significantly influenced by density variability. As illustrated in Fig.~\ref{fig:pitscheneder-height}, the surface deformation along the axis of symmetry throughout the entire melting process differs significantly between the cases with constant and variable densities. The surface topography is visualized as an iso-surface at the value $\alpha = 0.5$ in Fig.~\ref{fig:Pitscheneder_melting_topo} at $t = \SI{6}{\s}$. A drastic difference between incompressible and compressible cases is observed for the both sulfur concentrations in Fig.~\ref{fig:Pitscheneder_melting_topo}. The scenario incorporating variable density qualitatively agrees with the experimental data reported in the original study by Pitscheneder et al.~\cite{Pitscheneder1996} (Fig.~\ref{fig:Pitscheneder_150_pools_comparison}).
\revA{This agreement stems from the correct physical description of density changes affecting the melt volume, thermocapillary forces dominating convective transport in the melt pool, and the capillary forces stabilizing the gas--melt interface.}

The results presented in this \revALL{sub}section demonstrate that the proposed methodology for surface energy deposition processes accurately predicts evolution of the free surface during both melting and solidification. Consequently, this approach can be used to explain the physical mechanisms governing topography formation \cite[e.g.,][]{Kou2011, Scheller2001}.
Furthermore, the continuous melting and solidification process under energy deposition, as described in Ref.~\cite{Temmler2012structuring}, is expected to be more accurately simulated using the proposed methodology.

\section{Conclusions}
\label{sec:conclusions}
The present study introduces a comprehensive mathematical model and its numerical implementation to address the challenges of modeling metal melting due to surface energy deposition from an incident beam. A central focus of this work is on considering the effects of  variable liquid metal density on melt-pool dynamics and surface topology after solidification, along with providing a detailed discussion of the results and limitations of the proposed approach.

The developed mathematical model accounts for density variability and other key physical phenomena such as isothermal  transitions between liquid and solid phases, capillary and thermocapillary effects, and beam--surface interaction. Density variability is limited to thermal and phase transition dilatation mechanisms, without considering the effect of mechanical compressibility. The formulation is applicable to scenarios with either a free surface or open boundaries, \revALL{which accommodates thermal expansion}.

In addition to formulating the mathematical framework, two important issues related to the numerical implementation of the proposed model are considered. The first involves the stepwise movement of the solid--liquid interface due to the discretization of the liquid volume fraction in the enthalpy formulation. To address this challenge, a smoothing procedure for the mushy zone is proposed, where the liquid volume fraction is defined by a continuous and differentiable function. The second issue is the deterioration of global metal mass conservation. To ensure the global mass conservation without direct mass injection, a correction procedure based on modification of the dilatation rate equation during the pressure--velocity coupling step is introduced. These and other aspects of the developed mathematical model are implemented in OpenFOAM, enhancing the \texttt{interIsoFoam} solver.

The developed mathematical model and its numerical implementation have been validated against benchmark problems addressing various physical effects. In the one-dimensional Stefan problem, the relation between the melt front position and the generated flow in the liquid phase due to phase transition is clearly demonstrated. A differentiable sigmoid-based function is applied to evaluate the molten metal volume fraction, effectively mitigating the artificial stepwise movement of the solid--liquid interface, as shown in the Stefan problem. The issue of global mass-conservation loss and the effectiveness of the proposed mass-correction procedure are illustrated in the horizontal solidification problem with a free interface. The importance of accounting for both phase and thermal dilatation when predicting surface topography after solidification is demonstrated in the vertical solidification problem, where the influence of the sigmoid-based mushy zone smoothing on the position of the solidified metal surface is also highlighted. The final benchmark problem, involving the axisymmetric melting and solidification of metal from an incident laser beam, incorporates all the physical phenomena described by the mathematical model. This comprehensive case reveals only a minor influence of the variable metal density on the geometry of the melt pool during the melting process, while demonstrating a drastic impact on the topography of the solidified metal surface.

\revALL{The presented framework can be employed to investigate the physical processes in industrial technologies involving conduction-dominated melting---such as laser polishing, laser structuring by remelting, conduction-mode laser welding---where an accurate simulation of post-solidification surface topography and melt-pool dynamics is essential. Incorporating evaporation effects would extend the model to evaporation-driven regimes, enabling applications to additive manufacturing and keyhole welding.}

\begin{acknowledgments}
    Oleg A.~Rogozin thanks Oerlikon AM for its support in 2018--2021 years, when the research was at an early stage.
    The authors acknowledge the Computational Engineering Laboratory at Skolkovo Institute of Science and Technology and Shared Research Facilities ``High Performance Computing and Big Data'' of FRC CSC RAS (Moscow) for providing the computational resources that contributed to the research results reported in this paper.
\end{acknowledgments}

\bibliography{biblio}

\end{document}